\documentclass[10pt,aps,preprintnumbers,prd,noshowpacs,nofootinbib,noshowkeys,floatfix,superscriptaddress]{revtex4}
\usepackage[dvips]{graphics,graphicx}
\usepackage[colorlinks=true,linktocpage=true,linkcolor=blue,citecolor=blue]{hyperref}
\usepackage[usenames,dvipsnames]{color}
\usepackage{amsmath, amssymb,oldgerm}
\usepackage{multirow}
\usepackage{xcolor}
\usepackage[normalem]{ulem}  % \sout{old text} for strikeout
\usepackage{braket}
\usepackage{slashed}
\usepackage[mathscr]{eucal}
\usepackage[dvips]{graphics,graphicx}
 \usepackage{caption}
\usepackage{subcaption}

\newcommand{\n}{\nonumber}

\newcommand{\F}{\mathcal{F}}
\newcommand{\Pc}{\mathcal{P}}

\newcommand{\A}{\mathcal{A}}
\newcommand{\mC}{\mathfrak{C}}

\newcommand{\ms}{\mathfrak{s}}

\newcommand{\beq}{\begin{equation}}
\newcommand{\eeq}{\end{equation}}

\newcommand{\eqs}{Eqs.~}
\newcommand{\eq}{Eq.~}

\newcommand{\p}{\mathbf{p}}

\newcommand{\mG}{\mathcal{G}}
\newcommand{\mI}{\mathcal{I}}
\newcommand{\mL}{\mathcal{L}}
\newcommand{\msb}{s_\bot}

\begin{document}

\title{Polarization dynamics from moment equations}

\author{Nora Weickgenannt}

\affiliation{Institut de Physique Th\'eorique, Universit\'e Paris Saclay, CEA, CNRS, F-91191 Gif-sur-Yvette, France}

\author{Jean-Paul Blaizot}

\affiliation{Institut de Physique Th\'eorique, Universit\'e Paris Saclay, CEA, CNRS, F-91191 Gif-sur-Yvette, France}

\begin{abstract}
 We derive an expression for the local transverse polarization of a boost-invariant expanding system of massive particles, which involves a set of dynamical spin moments. Starting from spin kinetic theory, we obtain a closed set of equations of motion for these spin moments. These equations are valid during the full evolution of the system, from free streaming to local equilibrium, and can be used to study polarization phenomena in relativistic heavy-ion collisions.
\end{abstract}

\maketitle

\section{Introduction}

Since the first measurement of the polarization of Lambda hyperons in relativistic heavy-ion collisions~\cite{STAR:2017ckg}, polarization phenomena have been  actively investigated both on the experimental~\cite{Adam:2018ivw,ALICE:2019aid,STAR:2019erd} and theoretical~\cite{Becattini:2013vja,Becattini:2013fla,Becattini:2015ska,Becattini:2016gvu,Karpenko:2016jyx,Pang:2016igs,Xie:2017upb} sides. While it is commonly believed, and confirmed, that the global polarization along the orbital angular momentum of the system is mainly caused by the conversion of (thermal) vorticity into spin polarization~\cite{Liang:2004ph,Voloshin:2004ha,Betz:2007kg,Becattini:2007sr}, resembling the non-relativistic Barnett effect~\cite{Barnett:1935}, the origin of the momentum dependence of the local polarization is not yet well understood,  and is controversially discussed~\cite{Becattini:2017gcx,Florkowski:2019qdp,Florkowski:2019voj,Becattini:2020ngo,Liu:2021uhn,Fu:2021pok,Becattini:2021suc,Becattini:2021iol,Alzhrani:2022dpi,Jiang:2023vxp}. Since the models that assume the local polarization to be determined by thermal vorticity fail to describe the momentum dependence of the local polarization, alternative ideas have been proposed during the last years. In particular, the suggestion of including contributions from thermal shear to the local-equilibrium polarization appears to be a promising development~\cite{Liu:2021uhn,Fu:2021pok,Becattini:2021suc,Becattini:2021iol}. On the other hand, the effects of non-equilibrium spin dynamics on the polarization are still to be investigated.

In order to obtain a dynamical description of the polarization, lots of effort have been devoted to derive a theory of relativistic spin hydrodynamics~\cite{Florkowski:2017ruc,Florkowski:2017dyn,Florkowski:2018fap,Montenegro:2018bcf,Hattori:2019lfp,Bhadury:2020puc,Singh:2020rht,Montenegro:2020paq,Gallegos:2021bzp,Fukushima:2020ucl,Li:2020eon,Wang:2021ngp,Hu:2021pwh,Hongo:2021ona,Daher:2022xon,Weickgenannt:2022zxs,Weickgenannt:2022jes,Gallegos:2022jow,Cao:2022aku,Weickgenannt:2022qvh,Biswas:2023qsw,Weickgenannt:2023btk}. The main idea of spin hydrodynamics is to promote the spin tensor to a dynamical variable, in addition to the charge current and the energy-momentum tensor,  and to derive equations of motion for these quantities, e.g. starting from kinetic theory, and assuming the system to be in local equilibrium~\cite{Florkowski:2017ruc,Florkowski:2017dyn}, or close to local equilibrium~\cite{Weickgenannt:2022zxs,Weickgenannt:2022qvh,Weickgenannt:2023btk}. Since the forms of the energy-momentum tensor and of the spin tensor depend on the choice of a pseudo-gauge~\cite{Hehl:1976vr}, the evolution equations in such a theory suffer from an ambiguity related to the pseudo-gauge freedom~\cite{Speranza:2020ilk}. Although there have been suggestions for a reasonable choice of pseudo-gauge~\cite{Weickgenannt:2021cuo,Daher:2022xon,Weickgenannt:2022jes}, no consensus has been reached so far, and it appears desirable to derive equations of motion for the polarization which are  independent of the pseudo-gauge choice. Furthermore,  the early-time regime of relativistic heavy-ion collisions is not determined by local equilibrium, and the imprint of the dynamics at early time on the polarization measured at freeze-out has not been studied up to know. 

In this paper, we propose a way to calculate the polarization without referring to spin hydrodynamics. Our results are free of any pseudo-gauge ambiguity, and capture the dynamics of the system during the full evolution from free streaming to local equilibrium. We consider a boost-invariant system with Bjorken model~\cite{PhysRevD.27.140}, which is commonly used to describe the expansion of matter produced  in heavy-ion collisions. In the context of such a model, it has been found that the moment equations for the energy-momentum tensor without spin degrees of freedom feature an attractor solution, see, e.g., Refs.~\cite{Kurkela:2019set,Giacalone:2019ldn,Almaalol:2020rnu,Blaizot:2021cdv} for massless particles and Ref.~\cite{Jaiswal:2021uvv,Jaiswal:2022udf} for massive particles. Furthermore, it was shown that a simple truncation, combined with a suitable adjustment of  coefficients in these truncated moment equations needed in order to enforce the correct behavior both at the free-streaming and hydrodynamic fixed points, lead to an excellent description of the system at any time of the evolution~\cite{Blaizot:2021cdv}. These studies were extended to include chiral degrees of freedom in Ref.~\cite{Weickgenannt:2023nge}. In this paper, we employ a similar procedure to obtain equations of motion for various spin moments appearing in the polarization vector. For simplicity, we restrict ourselves to polarization in the transverse plane. Studies of the longitudinal polarization are left for future work. Furthermore, due to the translational invariance in Bjorken symmetry, the vorticity vanishes, and we hence automatically focus only on contributions to the polarization which do not emerge from vorticity.

Our efforts consist   first, in expressing the momentum dependent polarization vector in the particle rest frame in terms of the relevant dynamical spin moments, and second in deriving a closed set of equations of motion for the latter.
In order to achieve the first part, we expand the spin-dependent distribution function in spherical harmonics, and define corresponding spin moments as phase-space integrals of spherical harmonics weighted with the distribution function. Although the local polarization vector is a function of the three momentum of the particles, we are mainly interested in its dependence on the azimuthal momentum angle $\phi$. We may thus integrate over the absolute value and the polar angle of the three momentum, after which the polarization vector can be written as an infinite sum of spin moments. Then we derive the Boltzmann equation for the spin-dependent distribution function in Bjorken symmetry employing a relaxation time approximation. Assuming the collision term to be local, we impose microscopic conservation of spin angular momentum. This results in a matching condition which determines the equation of motion for the spin potential appearing in the local-equilibrium distribution function. We then derive an infinite system of coupled equations of motion for the spin moments from the Boltzmann equation. In the free-streaming limit, these equations of motion feature an unstable and a stable fixed point, where the moments decay with power laws. On the other hand, in the collision dominated regime, we distinguish moments which decay with power laws from those which decay exponentially. Given that the relaxation time is not extremely large and the system will have reached the hydrodynamic regime before freeze out, the exponentially decaying moments will be zero at the time any measurement takes place, regardless their dynamics at early time. Therefore, we drop these moments in the expression for the polarization vector. On the other hand, the spin moments which decay as power laws may be expanded in powers of relaxation time over proper time, $w^{-1}$. We find that higher moments start to contribute to the polarization vector at corresponding higher orders of $w^{-1}$. Making use of this result, we provide an expression for the polarization as a function of $\phi$, which is a power series of $w^{-1}$. The latter may be truncated at a given order $w^{-k}$, neglecting moments which decay at least $\sim w^{-(k+1)}$ faster than the local-equilibrium moments. In order to close the system of equations of motion for the dynamical spin moments, we approximate the higher moments, which are neglected in the polarization vector, by an interpolation between the free-streaming fixed point and the local-equilibrium regime. In this way, we take into account the dynamics of the system during the full time of the evolution. As an example, we give an explicit expression for the polarization using the next-to-leading-order truncation and provide the equations of motion for the spin moments appearing in this expression.

This paper is organized as follows. In Section \ref{polsec} we derive a general expression for the polarization vector as a function of $\phi$ in terms of spherical harmonics and spin moments. The Boltzmann equation for the spin-dependent distribution function in Bjorken symmetry is obtained in Section \ref{kinsec}. In Section \ref{spinpotsec} we discuss the spin potential and the matching conditions. The equations of motion for the spin moments are derived and analyzed in Section \ref{eomsec}. Finally, we present in Section \ref{resultsec} the main result of this work, consisting of the final expression for the polarization vector and the closed set of equations of motion. Conclusions are given in Section \ref{concsec}. Throughout this paper, we use the following notation and conventions, $a\cdot b \equiv a_\mu b^\mu$, $g_{\mu\nu}=\text{diag}(+,-,-,-)$, $\epsilon^{0123}=-\epsilon_{0123}=1$. We do not distinguish between upper and lower spatial indices of three vectors. The symbol $\ast$ indicates the complex conjugate.

\section{Polarization in heavy-ion collisions}
\label{polsec}

We consider the polarization vector~\cite{Becattini:2020sww,Weickgenannt:2022zxs}
\begin{equation}
    \Pi^\mu(\p) \equiv \frac{1}{2N} \int dS(\p) \int d\Sigma_\lambda p^\lambda \ms^\mu f(x,\p,\ms) \; ,
\end{equation}
where $\Sigma_\lambda$ is the freeze-out hypersurface, and $f(x,p,\ms)$ is the distribution function depending on the space-time position $x^\mu$, on the on-shell three-momentum $\p$, and on the spin four-vector $\ms^\mu$. Furthermore  $dS(\p)\equiv (m/\sqrt{3}\pi)d^4\ms\, \delta(p\cdot\ms)\delta(\ms^2+3)$ and is the integration measure on $\ms^\mu$, and $N$ is a normalization
\begin{equation}
    N\equiv \int d\Sigma_\lambda p^\lambda \int dS(\p)\, f\; .
\end{equation}
Since in heavy-ion collisions the polarization is measured in the particle rest frame~\cite{STAR:2017ckg}, we aim at calculating the following Lorentz transformed polarization three-vector,
\begin{equation}
    \boldsymbol{\Pi}_\star(\p)= \frac{1}{2N}\int dS(\p) \int d\Sigma_\lambda p^\lambda \left[ \boldsymbol{\ms}-\frac{(\boldsymbol{\ms}\cdot \p)\p}{E_p(m+E_p)} \right] f \; \label{pistar0}
\end{equation}
with $E_p\equiv \sqrt{\p^2+m^2}$.
In the following, we work up to first order in $\hbar$ and assume that polarization effects enter only at first order. Then, the distribution function takes the form~\cite{Weickgenannt:2020aaf}
\begin{equation}
    f(x,p,\ms)=\frac12 \left[\F(x,p) -\ms\cdot \A(x,p) \right] \label{ffa}
\end{equation}
with $\F(x,p)$ of zeroth order and $\A^\mu(x,p)$ of first order in $\hbar$. Since any component of $\A^\mu$ parallel to $p^\mu$ would vanish in the spin integration, we may assume $p\cdot\A=0$ without loss of generality.

For comparison to measurements of the local $\Lambda$-polarization, we need to access the dependence of $\boldsymbol{\Pi}_\star$ on the azimuthal momentum angle $\phi$~\cite{Becattini:2017gcx}. 
 In order to do so, it is convenient to expand the distribution function $f$ in terms of spherical harmonics as
\begin{equation}
  f(x,\p,\ms) = \sum_{n=0}^\infty \sum_{\ell=-n}^n N_{n\ell} \tilde{f}_{n\ell}(x,p,\ms) Y_n^\ell(\theta,\phi)\; , 
  \label{fspher}
\end{equation}
where
\begin{equation}
    Y_n^\ell(\theta,\phi)\equiv \Pc_n^\ell(\cos\theta) e^{i\ell\phi}\; ,
\end{equation}
$\Pc_n^\ell$ are associated Legendre polynomials, $\theta$ is the polar angle of the three-momentum $\p$, 
\begin{equation}
\cos\theta\equiv \frac{p_z}{p}\; , \qquad \qquad p\equiv \sqrt{\mathbf{p}^2}\; ,
\end{equation}
and $\phi$ is the azimuthal angle of $\p$. 
The coefficients $\tilde{f}_{n\ell}$ depend on $\p$ only through $p$. We also defined the normalization coefficients
\begin{equation}
N_{n\ell}\equiv \frac{2n+1}{4\pi}\frac{(n-\ell)!}{(n+\ell)!} \; \label{nnl}
\end{equation}
such that
\begin{equation}
  N_{n\ell} \int d\phi\, d\cos\theta\, Y^\ell_n(\theta,\phi) Y^k_m(\theta,\phi)=\delta_{nm} \delta_{\ell k}\; .
\end{equation}

In the following, we  consider the polarization for a system which depends on $x$ only through the proper time $\tau$, see the next section for details. We also assume that the freeze out takes place at constant proper time $\tau$. As we are interested in the dependence of $\boldsymbol{\Pi}_\star$ on $\phi$, we can perform the $p$ and $\theta$ integrations in the numerator and the denominator of \eq\eqref{pistar0}, respectively. Inserting \eq\eqref{fspher}, we obtain
\begin{align}
    \boldsymbol{\Pi}_\star(\phi)
        &= \frac{1}{2\mathcal{N}}\sum_{n=0}^\infty \sum_{\ell=-n}^n \int dp \int d\cos\theta\, p^2  E_p \left[ \mathbf{A}_{n\ell}-\frac{(\mathbf{A}_{n\ell}\cdot \p)\p}{E_p(m+E_p)} \right] N_{n\ell} Y_n^\ell(\theta,\phi)\; ,\label{pistar}
\end{align}
where we assumed that $N$ does not depend on $\phi$ and we defined
\begin{equation}
 \mathcal{N}(\tau)\equiv \frac{1}{\int d\Sigma_\tau} \int dp \int d\cos\theta\, p^2 N(\tau,p,\theta)  \; ,
\end{equation}
with $\Sigma_\tau$ the freeze-out hypersurface at constant $\tau$, as well as
\begin{equation}
 A^k_{n\ell}(\tau,p)\equiv\int dS(\p)\, \ms^k \tilde{f}_{n\ell}(\tau,p,\ms)=\int dS(\p)\, \int d\cos\theta \int d\phi\, \ms^k Y_n^\ell(\theta,\phi) f\; .  \label{aknell}
\end{equation}
Note that we made use of \eq\eqref{ffa} and of the relation $p\cdot \A=0$ to find that $\int dS(\p) \ms^k f= \A^k$, from which it follows that $\int dS(\p) \ms^k \tilde{f}_{n\ell}$ is a function of $\tau$ and $p$ only.
The integration over $d\cos\theta$ will make certain terms in \eq\eqref{pistar} vanish. For $\ell=0$ we have
\begin{equation}
   N_{n0} \int d\cos\theta\, \Pc_n^0(\cos\theta)=\delta_{n0}\; .
\end{equation}
Furthermore, we know that due to the symmetry of the associated Legendre polynomials
\begin{equation}
    \int d\cos\theta\, \Pc_n^\ell(\cos\theta)=0\; , \qquad \qquad n+\ell \text{ odd}\; .
\end{equation}
This means that the sum in \eq\eqref{pistar} needs to be taken only over the values of $\ell$ with $n+\ell$ even.

Note that the full dynamics of the polarization vector is contained in the coefficients $A^k_{n\ell}(\tau,p)$. We will now express these coefficients through spin moments and determine their equations of motion. To this end, we define the spin moments
\begin{align}
\mG_{n\ell r}^k&\equiv \int_{p\ms} \left(\frac{p}{E_p}\right)^r Y_n^\ell(\theta,\phi) \ms^k f\; ,\n\\
\mI_{n\ell r}^k&\equiv m  \int_{p\ms} \frac{1}{E_p} \left(\frac{p}{E_p}\right)^r Y_n^\ell(\theta,\phi) \ms^k f \; ,  \label{gidef} 
\end{align}
where we introduced the short-hand notation $\int_{p\ms}\equiv (m/\sqrt{3}\pi)\int d^3p\, d\ms_z\, d\theta_\ms$, with $\theta_\ms$ the polar angle of the spin three-vector in cylindrical coordinates. Note that $f$ in \eqs\eqref{gidef} should be evaluated for on-shell momentum and spin, i.e., $p^0\equiv E_p$, $\ms^0\equiv \p\cdot\boldsymbol{\ms}/E_p$, $\msb\equiv\sqrt{\ms^2_x+\ms_y^2}=\sqrt{\ms_0^2-\ms_z^2+3}$, such that $f$ depends only on $\p$, $\ms_z$, and $\theta_\ms$. The choice of cylindrical coordinates for the spin variable is made for convenience. Since $f$ is real and $\Pc_n^{-\ell}(\cos\theta)=(-1)^\ell [(n-\ell)!/(n+\ell)!]\Pc_n^\ell(\cos\theta)$ we have
\begin{equation}
\mG_{n(-\ell) r}^k= \int_{p\ms} \left(\frac{p}{E_p}\right)^r (-1)^\ell \frac{(n-\ell)!}{(n+\ell)!}\Pc_n^\ell(\cos\theta) e^{-i\ell\phi} \ms^k f  =  (-1)^\ell\frac{(n-\ell)!}{(n+\ell)!} \left(\mG_{n\ell r}^k\right)^\ast\; .
\label{gpm}
\end{equation}
Thus it is sufficient to calculate the moments with positive $\ell$ in order to determine the polarization.

Performing the $dp$-integration over the terms in the square brackets in \eq\eqref{pistar},
we obtain
\begin{equation}
2\int dp\, p^2 E_p A^k_{n\ell}= \mG^k_{n\ell0}\; ,   \label{pi1stterm}
\end{equation}
and, e.g., for $k=x$
\begin{align}
        -2\int dp\, p^2 E_p \frac{(\mathbf{A}_{n\ell}\cdot \p)p^x}{E_p(m+E_p)}&=-2\int dp\, p^2 \frac{1}{(m+E_p)}\int dS(p)\,  (\ms^x p^x+\ms^y p^y+\ms^z p^z) p^x  \tilde{f}_{n\ell}(x,p,\ms)\n\\
        &= \mathfrak{a}^+_{n\ell} \left(\mI^x_{n(\ell+2)0}-\mG^x_{n(\ell+2)0}+\mI^y_{n(\ell+2)0}-\mG^y_{n(\ell+2)0} \right)\n\\
        &+\mathfrak{b}^+_{n\ell} \left(\mI^x_{(n-2)(\ell+2)0}-\mG^x_{(n-2)(\ell+2)0}+\mI^y_{(n-2)(\ell+2)0}-\mG^y_{(n-2)(\ell+2)0} \right)\n\\
         &+\mathfrak{c}^+_{n\ell} \left(\mI^x_{(n+2)(\ell+2)0}-\mG^x_{(n+2)(\ell+2)0}+\mI^y_{(n+2)(\ell+2)0}-\mG^y_{(n+2)(\ell+2)0} \right)\n\\
     & + \mathfrak{a}^-_{n\ell} \left(\mI^x_{n(\ell-2)0}-\mG^x_{n(\ell-2)0}-\mI^y_{n(\ell-2)0}+\mG^y_{n(\ell-2)0} \right)\n\\
        &+\mathfrak{b}^-_{n\ell} \left(\mI^x_{(n-2)(\ell-2)0}-\mG^x_{(n-2)(\ell-2)0}+\mI^y_{(n-2)(\ell-2)0}-\mG^y_{(n-2)(\ell-2)0} \right)\n\\
         &+\mathfrak{c}^-_{n\ell} \left(\mI^x_{(n+2)(\ell-2)0}-\mG^x_{(n+2)(\ell-2)0}+\mI^y_{(n+2)(\ell-2)0}-\mG^y_{(n+2)(\ell-2)0} \right)\n\\  
         &+\mathfrak{a}^0_{n\ell} \left(\mI^x_{n\ell0}-\mG^x_{n\ell0}\right)+\mathfrak{b}^0_{n\ell} \left(\mI^x_{(n-2)\ell0}-\mG^x_{(n-2)\ell0}\right)+\mathfrak{c}^0_{n\ell} \left(\mI^x_{(n+2)\ell0}-\mG^x_{(n+2)\ell0}\right)\n\\
         &+\mathfrak{d}^+_{n\ell} \left(\mI^z_{n(\ell+1)0}-\mG^z_{n(\ell+1)0}\right)+\mathfrak{e}^+_{n\ell} \left(\mI^z_{(n-2)(\ell+1)0}-\mG^z_{(n-2)(\ell+1)0} \right)\n\\
         &+\mathfrak{f}^+_{n\ell} \left(\mI^z_{(n+2)(\ell+1)0}-\mG^z_{(n+2)(\ell+1)0} \right)+\mathfrak{d}^-_{n\ell} \left(\mI^z_{n(\ell-1)0}-\mG^z_{n(\ell-1)0}\right)\n\\
         &+\mathfrak{e}^-_{n\ell} \left(\mI^z_{(n-2)(\ell-1)0}-\mG^z_{(n-2)(\ell-1)0} \right)+\mathfrak{f}^-_{n\ell} \left(\mI^z_{(n+2)(\ell-1)0}-\mG^z_{(n+2)(\ell-1)0} \right)\; ,
         \label{pixlong}
\end{align}
where we made use of the properties of the associated Legendre polynomials to express the terms in the first line through spherical harmonics.
The coefficients are given in Appendix \ref{coeffapp}. 

Thus, in order to obtain the polarization, we need to determine all spin moments appearing in \eqs\eqref{pi1stterm} and \eqref{pixlong}. In the remainder of this paper, we will calculate the equations of motion for these spin moments, \eqs\eqref{gidef}, for a boost invariant expanding system. Making use of the properties of these equations of motion, we will then show how to truncate the sums in \eq\eqref{pistar} in a reasonable way, and provide closed equations of motion for the moments appearing in the final expression for the polarization vector.

\section{Spin kinetic theory for Bjorken expansion}
\label{kinsec}

We consider the Boltzmann equation~\cite{Weickgenannt:2020aaf}
\begin{equation}
p\cdot \partial f(x,p,\ms)=\mC[f]\;, \label{bbooll}
\end{equation}
where the collision term $\mC[f]$ is treated in the relaxation time approximation, viz.
\begin{equation}
\mC[f]=-p\cdot u \ \frac{f(x,p,\ms)-f_\text{eq}(x,p,\ms)}{\tau_R}.
\label{rta}
\end{equation}
Here $f_\text{eq}(x,p,\ms)$ is the local-equilibrium distribution function, $u^\mu$ the fluid velocity, and $\tau_R$ the relaxation time, which is assumed to be constant. We assume  Bjorken symmetry, i.e., boost invariance along the $z$-direction and translational invariance in the $x$-$y$-plane with $u^\mu\equiv (t,0,0,z)/\sqrt{t^2-z^2}$. 

Consider the distribution function $f(t,z,\p_\bot,p_z,\theta_\ms,\ms_z)$ with $\p_\bot\equiv(p_x,p_y)$. Boost invariance along the $z$-direction tells us that
\begin{equation}
f(t,z,p_z,\ms_z)=f(\tau,p_z^\prime,\ms_z^\prime)\; ,    
\end{equation}
where we suppressed the dependence on variables which are unaffected by the boost, and defined the proper time $\tau\equiv\sqrt{t^2-z^2}$, $p_z^\prime\equiv(p_z-E_pz/t)\gamma$, and $\ms_z^\prime\equiv (\ms_z-\ms_0 z/t)\gamma$ with $\gamma\equiv t/\tau$. We have therefore
\begin{align}
    \partial_z f(t,z,p_z,\ms_z)&= \partial_z f(\tau,p_z^\prime,\ms_z^\prime)|_{z=0}\n\\
    &= -\left(\frac{E_p}{\tau}\partial_{p_z}+\frac{\ms_0}{\tau}\partial_{\ms_z}\right) f(\tau,p_z,\ms_z)\;. 
\end{align}
Using this relation, one can rewrite the Boltzmann equation \eqref{bbooll} as follows
\begin{equation}
    \left(\partial_\tau-\frac{p_z}{\tau}\partial_{p_z}-\frac{p_z}{E_p^2}\frac{\mathbf{p}\cdot \boldsymbol{\ms}}{\tau}\partial_{\ms_z}\right)f(\tau,\p_\bot,p_z,\theta_\ms,\ms_z)=-\frac{f(\tau,\p_\bot,p_z,\theta_\ms,\ms_z)-f_\text{eq}(\tau,\p_\bot,p_z,\theta_\ms,\ms_z)}{\tau_R}\;. \label{boltzbjorkcomp}
\end{equation}

As shown in Appendix \ref{freeapp}, an exact analytical solution for \eq\eqref{boltzbjorkcomp} can be obtained under the assumption that the initial polarization at $\tau=\tau_0$ is restricted to the transverse plane, i.e., the initial distribution function depends on $\theta_\ms$ and $\ms_z$ only through $\ms_x\equiv \msb \cos\theta_\ms$ and $\ms_y\equiv \msb \sin\theta_\ms$, 
\begin{equation}
   f(\tau=\tau_0,\p_\bot,p_z,\theta_\ms,\ms_z)= f_\text{in}(\p_\bot,p_z,\ms_x,\ms_y)\; .
\end{equation}
With this, we find that
\begin{equation}
    f(\tau,p_\bot,p_z,\ms_z,\theta_\ms)=D(\tau,\tau_0) f_\text{in}\left(\p_\bot,p_z\frac{\tau}{\tau_0},\msb\cos\theta_\ms,\msb\sin\theta_\ms\right) + \int_{\tau_0}^\tau \frac{d\tau^\prime}{\tau_R} D(\tau,\tau^\prime) f_\text{eq}\left(\p_\bot,p_z\frac{\tau}{\tau_0},\msb\cos\theta_\ms,\msb\sin\theta_\ms\right) \label{easysol}
\end{equation}
with
\begin{equation}
    D(\tau_1,\tau_2)\equiv \exp\left(-\frac{\tau_1-\tau_2}{\tau_R}\right)
\end{equation}
solves \eq\eqref{boltzbjorkcomp} given that the equilibrium distribution function depends only on $\ms_x$ and $\ms_y$. We remark that this does not mean that it is independent of $\ms_z$, since $\ms_x$ and $\ms_y$ are functions of $\ms_z$ after integrating out the delta functions. Note that, although \eq\eqref{easysol} looks simple, solving it explicitly is actually involved, in part due the necessity to impose matching conditions to determine the parameters of $f_\text{eq}$ (see below). In the following, we will derive an alternative, easier way to obtain the polarization through equations of motion for the moments. We will restrict ourselves to a situation where both the initial polarization and equilibrium polarization are restricted to the transverse plane in order to keep the discussion compact. The case of general polarization will be investigated in the future. The distribution function \eqref{ffa} can then be written as
\begin{equation}
f(\tau,\p_\bot,p_z,\ms_z,\theta_\ms)=\frac12\left[\F(\tau,\p)+\msb \cos\theta_\ms\, \A^x(\tau,\p)+\msb \sin\theta_\ms\, \A^y(\tau,\p)\right]\; .
\label{ftransvpol}
\end{equation}

\section{Spin potential}
\label{spinpotsec}

In spin kinetic theory, the local-equilibrium distribution function is given by~\cite{Weickgenannt:2020aaf}
\begin{equation}
    f_\text{eq}=\frac{1}{(2\pi\hbar)^3}\exp\left(-\beta\cdot p + \frac{\hbar}{4} \Omega_{\mu\nu}\Sigma_\ms^{\mu\nu}\right)\; , \label{eqgen}
\end{equation}
where $\Sigma_\ms^{\mu\nu}\equiv -(1/m)\epsilon^{\mu\nu\alpha\beta}p_\alpha \ms_\beta$ is the dipole-moment tensor  and $\beta_\mu$ and $\Omega_{\mu\nu}$ are Lagrange multipliers. While $\beta_\mu\equiv \beta u_\mu$, equal to the ratio of the  fluid velocity to the temperature $T=1/\beta$, is related to the conservation of the four momentum, $\Omega_{\mu\nu}$ is the so-called spin potential and is related to the conservation of total angular momentum. Imposing Bjorken symmetry and a dependence on $\ms_x$ and $\ms_y$ only, one reduces \eq\eqref{eqgen}  to
\begin{equation}
    f_\text{eq} (\p_\bot,p_z,\msb\cos\theta_\ms,\msb\sin\theta_\ms)= \frac{1}{(2\pi\hbar)^3} \exp\left[-\beta E_p-\frac{\hbar}{2m} \kappa_0^z (p^x\sin\theta_\ms-p^y\cos\theta_\ms) \msb\right] \label{feqq}
\end{equation}
with $\kappa_0^z\equiv-\Omega^{z0}$ being the only nonzero component of $\Omega^{\mu\nu}$. The Lagrange multipliers $\beta_\mu$ and $\Omega_{\mu\nu}$ are determined through matching conditions. In this work, we consider a local collision term, i.e., no orbital angular momentum is transferred into spin and the dipole-moment tensor is conserved separately in collisions~\cite{Weickgenannt:2020aaf}. Since in this case the collision term \eqref{rta} has to respect the microscopic conservation of spin angular momentum in addition to the conservation of linear momentum,
\begin{align}
    \int d^3p\, dS(\p)\, p^\mu \mC[f]&=0\; ,\n\\
    \int d^3p\, dS(\p)\, \Sigma_\ms^{\mu\nu} \mC[f]&=0 \; ,
    \label{consmic}
\end{align}
we impose the matching conditions
\begin{align}
   \int d^3p\, dS(\p)\, p^\mu f&= \int d^3p\, dS(\p)\, p^\mu f_\text{eq} ,\n\\
    \int d^3p\, dS(\p)\, \Sigma_\ms^{\mu\nu} f&= \int d^3p\, dS(\p)\, \Sigma_\ms^{\mu\nu} f_\text{eq}\; .
    \label{matching}
\end{align}
Considering  $\mu=0$, $\nu=z$ in the second equation \eqref{matching} and using \eq\eqref{gidef}, we obtain
\begin{equation}
\text{Im}\, \mG^{x}_{111}-\text{Re}\, \mG^y_{111} = \text{Im}\, \mG^{x}_{111,\text{eq}}-\text{Re}\, \mG^y_{111,\text{eq}}\; .   \label{imgreg}
\end{equation}
Inserting \eq\eqref{feqq} on the right-hand side of \eq\eqref{imgreg}, we then find
\begin{equation}
    \kappa_0^z= -\frac{1}{2\hbar}\left(\int d^3p\, p_x^2 \F_\text{eq}\right)^{-1} \sigma\; , \label{kappa0z}
\end{equation}
where we defined
\begin{equation}
    \sigma\equiv \text{Im}\, \mG^{x}_{111}-\text{Re}\, \mG^y_{111}\; . \label{sigma}
\end{equation}
Thus, $\sigma$ is the only quantity that we need, in addition to $\mG^k_{n\ell0}$ and $\mI^k_{n\ell0}$, in order to obtain the polarization, since $\sigma$ determines the spin potential, and consequently the equilibrium values of the spin moments.

We remark that \eqs\eqref{consmic} are properties of the kinetic theory and are therefore independent of the concept of spin hydrodynamics. However, it is possible to relate \eqs\eqref{consmic} to macroscopic conservation laws for the energy-momentum tensor $T^{\mu\nu}$ and the total angular-momentum tensor $J^{\lambda,\mu\nu}$, respectively, 
\begin{align}
    \partial_\mu T^{\mu\nu}&=0\; , & \partial_\lambda J^{\lambda,\mu\nu}&=0\; .
    \label{conslaws}
\end{align}
In the framework of spin hydrodynamics, the total angular momentum is split into orbital and spin parts,
\begin{equation}
    J^{\lambda,\mu\nu}= x^\mu T^{\lambda\nu}-x^\nu T^{\lambda\mu}+\hbar S^{\lambda,\mu\nu}\, ,
    \label{split}
\end{equation}
$S^{\lambda,\mu\nu}$ being the spin tensor. The choice of the splitting \eqref{split},  and therefore that of the energy-momentum tensor and of the spin tensor, depends on the pseudo-gauge, the conservation laws \eqref{conslaws} being independent of the latter~\cite{Hehl:1976vr,Speranza:2020ilk}. In the  Hilgevoord-Whouthuysen (HW) pseudo-gauge~\cite{HILGEVOORD19631} the spin tensor is conserved separately, as long as only local collisions are considered~\cite{Weickgenannt:2020aaf}. The HW energy-momentum tensor reads~\cite{Speranza:2020ilk}
\begin{equation}
 T^{\mu\nu}=  \frac14 \int_{p\ms} \frac{1}{E_p^2} p^\mu p^\nu f \label{emtensor} 
\end{equation}
and the HW spin tensor is given by~\cite{Speranza:2020ilk}
\begin{equation}
    S^{\lambda,\mu\nu}= \frac18\int_{p\ms} \frac{1}{E_p^2} p^\lambda \Sigma_\ms^{\mu\nu} f-\frac{\hbar}{4m^2} \partial^{[\nu} T^{\mu]\lambda}
    \label{spintensor}
\end{equation}
with the conservation equation
\begin{equation}
    \partial_\lambda S^{\lambda,\mu\nu}=0\; .
\end{equation}
Comparing the right-hand sides of \eqs\eqref{emtensor} and \eqref{spintensor} with \eqs\eqref{matching}, we see that the latter are equivalent to  the matching conditions
\begin{align}
    u_\mu T^{\mu\nu}&=u_\mu T^{\mu\nu}_\text{eq}\; , \n\\
    u_\lambda S^{\lambda,\mu\nu}&= u_\lambda S^{\lambda,\mu\nu}_\text{eq}\; ,
    \label{matching2}
\end{align}
and ensure the conservation laws \eqref{conslaws}. Equations of motion for the spin tensor may be obtained by expressing the first term in \eq\eqref{spintensor} in terms of the spin moments $\mG^k_{n\ell r}$ in \eq\eqref{gidef}. However, for our purpose of calculating the polarization, the spin tensor itself is not needed. Instead, we will focus on evaluating the spin moments appearing in \eqs\eqref{pi1stterm} and \eqref{pixlong}.

% and
% \begin{equation}
%     K\equiv -\frac{1}{2\hbar}\left(\int d^3p\, p_x^2 \F_\text{eq}\right)^{-1}%=\frac{3}{4\hbar}\left(\int d^3p\, p^2 \F_\text{eq}\right)^{-1}
% \end{equation}

\section{Equations of motion for the spin moments}
\label{eomsec}

Making use of the Boltzmann equation \eqref{boltzbjorkcomp}, the properties of associated Legendre polynomials and the identity \eqref{dsx0}, we obtain the equations of motion
\begin{align}
    \partial_\tau \mG^x_{n\ell r}
    =& -\frac{1}{\tau} \left( {a}_{n\ell r} \mG^x_{n\ell r}+ {b}_{n\ell r} \mG^x_{(n-2)\ell r}+{c}_{n\ell r} \mG^x_{(n+2)\ell r}+{d}_{n\ell r} \mG^x_{n\ell(r+2)}+{e}_{n\ell r} \mG^x_{(n-2)\ell(r+2)}+{f}_{n\ell r} \mG^x_{(n+2)\ell(r+2)} \right)\n\\
    &-\frac{1}{\tau_R} \left(\mG^x_{n\ell r}-\mG^x_{n\ell r,\text{eq}}\right)\; ,
    \label{geom}
\end{align}
where the coefficients and the detailed calculation are shown in Appendix \ref{peomapp}. The equations of motion for $\mI^x_{n\ell r}$ are obtained analogously as
\begin{align}
    \partial_\tau \mI^x_{n\ell r}
    =& -\frac{1}{\tau} \left( {a}_{n\ell r} \mI^x_{n\ell r}+ {b}_{n\ell r} \mI^x_{(n-2)\ell r}+{c}_{n\ell r} \mI^x_{(n+2)\ell r}+\tilde{d}_{n\ell r} \mI^x_{n\ell(r+2)}+\tilde{e}_{n\ell r} \mI^x_{(n-2)\ell(r+2)}+\tilde{f}_{n\ell r} \mI^x_{(n+2)\ell(r+2)} \right)\n\\
    &-\frac{1}{\tau_R} \left(\mI^x_{n\ell r}-\mI^x_{n\ell r,\text{eq}}\right)\; ,
    \label{ieom}
\end{align}
where the coefficients in the first three terms are identical to those in \eq\eqref{geom}, and the last three coefficients can be found in Appendix \ref{peomapp}. The equations of motion for $\mG^y_{n\ell r}$ and $\mI^y_{n\ell r}$ are trivially obtained from \eqs\eqref{geom} and \eqref{ieom} by substituting $x\mapsto y$. In contrast to the massless case, where each equation of motion for the orthogonal moments couples only to two other moment~\cite{Weickgenannt:2023nge}, in the massive case the use of an orthogonal basis leads to an additional coupling of moments with different $r$. Note however that we need only moments with $r=0$ for the polarization vector, see \eqs\eqref{pi1stterm} and \eqref{pixlong}. In order to solve the systems of equations of motion \eqref{geom} and \eqref{ieom} for a finite number of spin moments, one requires a reasonable truncation which closes the sets of equations. In the following two subsections, we will discuss the equations of motion in the free-streaming regime and in the hydrodynamic regime, respectively. Making use of the results obtained from this discussion, we will then be able to close the equations of motion.

\subsection{Free-streaming regime}

The free-streaming regime of the equations of motion for the spin moments, i.e. the regime $\tau\ll\tau_R$, is discussed in Appendix \ref{ffpapp}. In particular,
we show there that,  as long as collisional effects can be neglected, the system  approaches the stable free-streaming fixed point where the spin moments are related through
\begin{align}
   \frac{\mG^x_{n\ell r}}{\mG^x_{mkr}}&=\frac{\Pc_n^\ell(0)}{\Pc_m^k(0)}\; , & n+\ell \text{ even}\; ,\n\\
   \frac{\mG^x_{n\ell r}}{\mG^x_{mkr}}&=\frac{(n+\ell)\Pc_{n-1}^\ell(0)}{(m+k)\Pc_{m-1}^k(0)}\; , & n+\ell \text{ odd} \; .
   \label{fsfp}
\end{align}
Using these relations, it is possible to calculate the polarization at the free-streaming fixed point, see Appendix \ref{freepolapp}. When collisional effects start to play a role, the system leaves the free-streaming fixed point  and enters the hydrodynamic regime. We expect that any measurement of the polarization  takes place when the system has reached this regime. Hence, we need an expression for the polarization  valid for $\tau\gg\tau_R$, which we construct in the next subsection.

\subsection{Hydrodynamic regime}

In the hydrodynamic regime, $\tau\gg\tau_R$, the collision term, given respectively by the last lines of \eqs\eqref{geom} and \eqref{ieom}, determines the dynamics of the system. Inserting \eq\eqref{feqq} into \eqs\eqref{gidef}, we find that the local-equilibrium spin moments appearing in the collision term are given by
\begin{align}
    \mG^x_{n\ell r,\text{eq}}&=\frac{2\hbar}{m}\int d^3p\, Y_n^\ell(\theta,\phi) E_p \left(\frac{p}{E_p}\right)^r \kappa_0^z\, p \sin\theta \sin \phi\, \F\n\\
    &=-\frac{2i\hbar}{m}\int d^3p\, Y_n^\ell(\theta,\phi) E_p \left(\frac{p}{E_p}\right)^r \kappa_0^z\, p \left(\frac12Y_1^1(\theta,\phi)-Y_1^{-1}(\theta,\phi)\right)\F\n\\
    &= \delta_{n1}\delta_{\ell1} \mG^x_{11 r,\text{eq}}+\delta_{n1} \delta_{\ell(-1)}\mG^x_{1(-1)r,\text{eq}}\; .
\end{align}
Note that $\left(\mG^x_{1(-1) r,\text{eq}}\right)^\ast=-(1/2)\mG^x_{11 r,\text{eq}}$.
The same holds for $\mI^x_{n\ell r}$. We also obtain
\begin{align}
    \mG^y_{n\ell r,\text{eq}}&=\frac{2\hbar}{m}\int d^3p\, Y_n^\ell(\theta,\phi) E_p \left(\frac{p}{E_p}\right)^r \kappa_0^z\, p \sin\theta \cos \phi\, \F\n\\
    &=\frac{2\hbar}{m}\int d^3p\, Y_n^\ell(\theta,\phi) E_p \left(\frac{p}{E_p}\right)^r \kappa_0^z\, p \left(\frac12Y_1^1(\theta,\phi)+Y_1^{-1}(\theta,\phi)\right)\F\n\\
    &= \delta_{n1}\delta_{\ell1} \mG^y_{11 r,\text{eq}}+\delta_{n1} \delta_{\ell(-1)}\mG^y_{1(-1)r,\text{eq}}\; ,
\end{align}
and analogously for $\mI^y_{n\ell r}$.
Inserting these results on the right-hand sides of \eqs\eqref{geom} and \eqref{ieom}, we find that only the moments with $|\ell|=1$ and $n$ odd couple to equilibrium quantities in the equations of motion. All other equations of motion do not contain equilibrium contributions, and therefore become for $\tau\gg\tau_R$, e.g.,
\begin{equation}
    \partial_\tau \mG^x_{n\ell r}=-\frac{1}{\tau_R}\mG_{n\ell r}^x\; .
\end{equation}
Thus, these moments decay exponentially in the hydrodynamic regime
\begin{equation}
    \mG^x_{n\ell r}\sim e^{-\tau/\tau_R}\; , \qquad \qquad |\ell|\neq 1 \text{ or } n \text{ even}\; ,
\end{equation}
and do not feature a power expansion around local equilibrium.

On the other hand, the moments $\mG^x_{n(\pm1)r}$ with $n$ odd couple to $\mG^x_{1(\pm1)r}$, which is nonzero in local equilibrium. These moments decay therefore with power laws, and it is possible to expand them around local equilibrium
\begin{equation}
\mG^x_{n1r}=\mG^x_{11r,\text{eq}} \sum_{k=q_n}^\infty g^{x(k)}_{nr} w^{-k} \label{expandgle}
\end{equation}
with $w\equiv \tau/\tau_R$ and dimensionless parameters $g^{x(k)}_{nr}$. From the equations of motion \eqref{geom} we find that there are no contributions to the sum smaller than $k=(n-1)/2$, therefore we can set $q_n=(n-1)/2$ without loss of generality. This can be proven as follows. For $n=1$ the statement is trivial. Furthermore, we know that $q_n>0$ for $n>1$, which validates the statement for $n=3$. Assuming $q_n=(n-1)/2$ for $n-2$, we obtain for $n\geq3$
\begin{align}
    -\sum_{k=q_n}^\infty z_{nr}^{(k)} k w^{-k-1} =& - \left( \bar{a}_{n\ell r} \sum_{k=q_n}^\infty z_{nr}^{(k)}  w^{-k-1}+ \bar{b}_{n\ell r} \sum_{k=(n-3)/2}^\infty z_{(n-2)r}^{(k)}  w^{-k-1}+\bar{c}_{n\ell r} \sum_{k=q_{n+2}}^\infty z_{(n+2)r}^{(k)}  w^{-k-1}\right.\n\\
    &\left.+\bar{d}_{n\ell r} \sum_{k=q_n}^\infty z_{n(r+2)}^{(k)}  w^{-k-1}+\bar{e}_{n\ell r} \sum_{k=(n-3)/2}^\infty z_{(n-2)(r+2)}^{(k)}  w^{-k-1}+\bar{f}_{n\ell r} \sum_{k=q_{n+2}}^\infty z_{(n+2)(r+2)}^{(k)}  w^{-k-1} \right)\n\\
    &-\sum_{k=q_n}^\infty z_{nr}^{(k)}  w^{-k}\; . \label{expandgineom}
\end{align}
For $n>3$, there is no contribution $\sim w^{-1}$ from the terms in the brackets. Thus $q_n>1$ for $n>3$. Inserting this result back into \eq\eqref{expandgineom}, we find with the same reasoning that $q_n>2$ for $n>5$. This can be continued to any $n$, thus $q_n=(n-1)/2$. The same reasoning applies also for $\mG^y_{n\ell r}$, $\mI^x_{n\ell r}$, and $\mI^y_{n\ell r}$, hence, e.g.,
\begin{equation}
\mI^x_{n1r}=\mI^x_{11r,\text{eq}} \sum_{k=q_n}^\infty i^{x(k)}_{nr} w^{-k}\; . \label{expandile}
\end{equation}

\section{Polarization vector from closed moment equations}
\label{resultsec}

The exact determination of the polarization \eqref{pistar} requires, in principle, an infinite number of spin moments, see \eqs\eqref{pi1stterm} and \eqref{pixlong}. However, taking into account the results of the previous section, we may simplify \eq\eqref{pistar} by neglecting all spin moments which decay exponentially in the hydrodynamic regime. In heavy-ion collisions, the polarization is measured at freeze out, where the system is generally supposed to be already close to local equilibrium. Therefore, all exponentially decaying spin moments will have disappeared at the time of the measurement, regardless their initial conditions or dynamics in the free-streaming regime. 
Assuming that the system is close enough to local equilibrium that the power expansion \eqref{expandgle} converges and the exponentially decaying moments have disappeared, we obtain by inserting \eqs\eqref{pi1stterm} and \eqref{pixlong} with \eqs\eqref{expandgle} and \eqref{expandile} into \eq\eqref{pistar} for $k=x$
\begin{align}
    \Pi_\star^x(\phi)
    &= \frac{1}{\mathcal{N}} \, \text{Re}\, \Bigg\{ \sum_{k=0}^\infty \sum_{n=1,3,\ldots,2k+1} \left[\left(h_n\mG^x_{110,\text{eq}}  g_{n1}^{x(k)}+\tilde{h}_n\mI^x_{110,\text{eq}} i_{n1}^{x(k)}\right)e^{i\phi}\right.\n\\
    &\left. -\bar{h}_n\left(\mG^x_{110,\text{eq}}  g_{n1}^{x(k)}-\mI^x_{110,\text{eq}} i_{n1}^{x(k)}+\mG^y_{110,\text{eq}} g_{n1}^{y(k)}-\mI^y_{110,\text{eq}} i_{n1}^{y(k)}\right) e^{-i\phi} \right] w^{-k} \Bigg\}\n\\
    &-\frac{1}{\mathcal{N}}  \, \text{Re}\, \Bigg[\sum_{k=1}^\infty \sum_{n=3,5,\ldots,2k+1} \hat{q}_n\left(\mG^x_{110,\text{eq}}  g_{n1}^{x(k)}-\mI^x_{110,\text{eq}} i_{n1}^{x(k)}+\mG^y_{110,\text{eq}} g_{n1}^{y(k)}-\mI^y_{110,\text{eq}} i_{n1}^{y(k)}\right)w^{-k} e^{3i\phi}\Bigg]\; . \label{pixnle}
\end{align}
In order to derive \eq\eqref{pixnle}, we used \eq\eqref{gpm}, $N_{n(-\ell)}=[(n+\ell)!/(n-\ell)!]^2 N_{n\ell}$, which follows from \eq\eqref{nnl}, and various relations between the coefficients \eqref{coeffhz} for positive and negative $\ell$, which can be derived from \eqs\eqref{abcpol}. The calculation and the factors are given in Appendix \ref{coeffapp}.

Although \eq\eqref{pixnle} still looks a bit complicated, the dependence on $\phi$ is actually simple. We also remark that all terms are zero if integrated over $\phi$, i.e., the global polarization in that case vanishes. The convergence of the power series in \eq\eqref{pixnle} is controlled by the terms $w^{-k}$. In practice, we want to obtain the polarization at a given, sufficiently large value of $w$, the freeze-out time, and one can truncate the series at some small $k$, neglecting moments which decay at least $\sim w^{-(k+1)}$ faster than the equilibrium quantities. The approximation is systematically improvable by using higher maximal values for $k$. At any truncation, we may re-express the $k$-dependent quantities through \eqs\eqref{expandgle} or \eqref{expandile} in terms of the full spin moments. Then, we will deal with a limited number of dynamical moments, which can be calculated from the equations of motion \eqref{geom} and the analogues for the other moments. We remark that the expansion in \eq\eqref{pixnle} is not equivalent to a gradient expansion of the Boltzmann equation itself around local equilibrium such as the Chapman-Enskog expansion. For the latter, only the equilibrium potentials (which in our case would be only $\beta$ and $\kappa_0^z$) are dynamical, and all other quantities are expressed as their gradients. On the other hand, in our approach the expansion in $w^{-1}$ only separates faster decaying moments from slower ones, taking into account the dynamics of the latter during the full evolution of the system. For our purpose, it is therefore not sufficient to consider only the hydrodynamic regime of the equations of motion, since the dynamics at early time may leave an imprint on the moments which are not negligible at freeze out.
Instead, following Ref.~\cite{Blaizot:2021cdv}, we assume that the system evolves from the free-streaming fixed point to the hydrodynamic regime, i.e., the system reaches the stable free-streaming fixed point before the collisions become relevant for the dynamics. 
Then, in order to close equations of motion \eqref{geom} and \eqref{ieom}, which determine the dynamical spin moments, at $n=n_\text{max}\equiv 2k_\text{max}+1$ with $k_\text{max}>0$, we may replace
\begin{equation}
\mG^x_{(n_\text{max}+2)10}\rightarrow \frac{\Pc_{n_\text{max}+2}^1(0)}{\Pc_{n_\text{max}}^1(0)} \mG_{n_\text{max}10}^x\; ,
\label{replacegmax}
\end{equation}
which is valid both at the free-streaming fixed point due to \eqs\eqref{fsfp} and in local equilibrium, since both moments vanish at the latter. 
Analogously, remembering the second identity in \eqs\eqref{stabfprel}, we may ignore the terms which couple moments of different $r$ as long as we are close to the stable free-streaming fixed point. For the moments which vanish in local equilibrium, we can drop the terms proportional to $d_{n\ell r}$, $e_{n\ell r}$, and $f_{n\ell r}$ during the full evolution. On the other hand, if these terms contain moments which are nonzero in local equilibrium, we need an interpolation to connect the two regimes, c.f.\ Ref.~\cite{Jaiswal:2022udf}. Since the sum of the last three terms vanishes at the free-streaming fixed point, only the contributions from the hydrodynamic regime need to be taken into account. We use the following interpolation between the two regimes
\begin{equation}
    \mG^x_{n12}\rightarrow -e^{-w/2} \Pc_n^1(0) \mG^x_{112}+\left(1-e^{-w/2}\right)\mG^x_{n12,\text{eq}}\; . \label{r2intpol}
\end{equation}
Here the first term corresponds to the free-streaming fixed point and gives the major contribution for $w\ll1$, while the second term ensures the spin moment to approach their local-equilibrium values for $w\gg1$. Note that we used $\Pc_1^1(0)=-1$ in the first term. We chose to express all 
 spin moments $\mG^x_{n12}$ as a function of $\mG^x_{112}$ for convenience, choosing any other value for $n$ with the corresponding ratio of associated Legendre polynomials in the first term would lead to the same result. Applying the replacements \eqref{replacegmax} and \eqref{r2intpol} in \eq\eqref{geom}, we obtain the following closed set of equations of motion
\begin{align}
    \partial_w \mG^x_{n10}&= -\frac1w \left( {a}_{n10} \mG^x_{n10}+ {b}_{n10} \mG^x_{(n-2)10}+{c}_{n10} \mG^x_{(n+2)10}\right)-\frac1w\left(1-e^{-w/2}\right)\left(\delta_{n1}{d}_{n10} +\delta_{n3}{e}_{n1 0} \right) \mG^x_{112,\text{eq}}\n\\
    &-\left(\mG^x_{n10}-\delta_{n1}\mG^x_{110,\text{eq}} \right)
    \label{eomgfin1}
\end{align}
for $n<n_\text{max}$ and
\begin{align}
    \partial_w \mG^x_{n10}&= -\frac1w \left[ \left({a}_{n10}+{c}_{n10}\frac{\Pc_{n+2}^1(0)}{\Pc_{n}^1(0)}\right) \mG^x_{n10}+ {b}_{n10} \mG^x_{(n-2)10}\right]+\frac1w\left(1-e^{-w/2}\right)\delta_{n3}{e}_{n\ell 0}  \mG^x_{112,\text{eq}}-\mG^x_{n10} \label{eomgfin2}
\end{align}
for $n=n_\text{max}$. The closed equations of motion for $\mI^x_{n\ell 0}$ can be obtained analogously from \eq\eqref{ieom} using the respective coefficients $\tilde{d}_{n\ell 0}$, $\tilde{e}_{n\ell0}$, and $\tilde{f}_{n\ell0}$. Furthermore, the equations of motion for the $y$-components are identical to those for the $x$-components with $x\mapsto y$.

In order to solve \eqs\eqref{eomgfin1} and \eqref{eomgfin2}, we need to determine $\mG^x_{110,\text{eq}}$ and $\mG^x_{112,\text{eq}}$, which depend on $\beta$ and $\kappa_0^z$ given by \eq\eqref{kappa0z}. The equation of motion for $\beta$ is determined from the first condition in \eqs\eqref{matching} and the equation of motion for
\begin{equation}
  T^{00}\equiv  \varepsilon= \frac{2}{(2\pi\hbar)^3}\int d^3p\, E_p\,  e^{-\beta E_p} \; ,
  \label{epsbeta}
\end{equation}
shown in Appendix \ref{emtensapp}.
Once $\beta$ is known, $\mG^x_{110,\text{eq}}$ and $\mG^x_{112,\text{eq}}$ are easiest obtained from $\sigma$ in \eq\eqref{sigma} through
\begin{align}
 \mG^x_{110,\text{eq}} &=  \frac12\left(\int d^3p\, E_p p\,  e^{-\beta E_p}\right) \left(\int d^3p\,  p^2 e^{-\beta E_p}\right)^{-1} \sigma\; ,\n\\
 \mG^x_{112,\text{eq}} &=  \frac12\left(\int d^3p\, \frac{p^3}{E_p}\,  e^{-\beta E_p}\right) \left(\int d^3p\,  p^2 e^{-\beta E_p}\right)^{-1} \sigma \; ,\n\\
 \mI^x_{110,\text{eq}} &=  \frac{m}{2}\left(\int d^3p\,  p\,  e^{-\beta E_p}\right) \left(\int d^3p\,  p^2 e^{-\beta E_p}\right)^{-1} \sigma\; ,\n\\
 \mI^x_{112,\text{eq}} &=  \frac{m}{2}\left(\int d^3p\, \frac{p^3}{E_p^2}\,  e^{-\beta E_p}\right) \left(\int d^3p\,  p^2 e^{-\beta E_p}\right)^{-1} \sigma \; ,
 \label{eqmom}
\end{align}
where we used 
\begin{equation}
    \text{Im}\, \mG^{x}_{111}=-\text{Re}\, \mG^y_{111}\; .
\end{equation}
The equation of motion for $\sigma$ can be obtained from its definition \eqref{sigma} and \eq\eqref{geom} for $n=\ell=r=1$. Noting that $a_{111}=1$ and $c_{111}=d_{111}=f_{111}=0$, see \eq\eqref{abcdefbar}, and using \eq\eqref{imgreg}, we obtain the simple equation of motion
\begin{equation}
    \partial_w \sigma = -\frac1w \sigma \; . \label{sigmaeom}
\end{equation}
 
As an example, we show the expression for $\Pi^x_\star(\phi)$ from \eq\eqref{pixnle} including moments which decay at most $\sim w^{-1}$ faster than the equilibrium moments,
\begin{align}
    \Pi_\star^x(\phi)
    &= \frac{1}{\mathcal{N}} \, \text{Re}\, \Bigg\{  \sum_{n=1,3} \left[\left(h_n\mG^x_{n10}  +\tilde{h}_n\mI^x_{n10} \right)e^{i\phi} -\bar{h}_n\left(\mG^x_{n10}  -\mI^x_{n10} +\mG^y_{n10}-\mI^y_{n10}\right) e^{-i\phi} \right]\n\\
    &-\hat{q}_3\left(\mG^x_{310}  -\mI^x_{310} +\mG^y_{310}-\mI^y_{310}\right) e^{3i\phi}\Bigg\}\; .\label{pixnle1st}
\end{align}
The relevant equations of motion are
\begin{align}
    \partial_w \mG^x_{110}&= -\frac1w \left( {a}_{110} \mG^x_{110}+{c}_{110} \mG^x_{310}\right)-\frac1w\left(1-e^{-w/2}\right){d}_{110} \mG^x_{112,\text{eq}}-\left(\mG^x_{110}-\mG^x_{110,\text{eq}} \right)\; ,\n\\
    \partial_w \mG^x_{310}&= -\frac1w \left[ \left({a}_{310}+{c}_{310}\frac{\Pc_{5}^1(0)}{\Pc_{3}^1(0)}\right) \mG^x_{310}+ {b}_{310} \mG^x_{110}\right]+\frac1w\left(1-e^{-w/2}\right){e}_{31 0}  \mG^x_{112,\text{eq}}-\mG^x_{310} \; ,\n\\
     \partial_w \mI^x_{110}&= -\frac1w \left( {a}_{110} \mI^x_{110}+{c}_{110} \mI^x_{310}\right)-\frac1w\left(1-e^{-w/2}\right)\tilde{d}_{110} \mI^x_{112,\text{eq}}-\left(\mI^x_{110}-\mI^x_{110,\text{eq}} \right)\; ,\n\\
    \partial_w \mI^x_{310}&= -\frac1w \left[ \left({a}_{310}+{c}_{310}\frac{\Pc_{5}^1(0)}{\Pc_{3}^1(0)}\right) \mI^x_{310}+ {b}_{310} \mI^x_{110}\right]+\frac1w\left(1-e^{-w/2}\right)\tilde{e}_{31 0}  \mI^x_{112,\text{eq}}-\mI^x_{310} \; .
\end{align}
To solve these equations, one also needs to determine the appearing equilibrium spin moments. Using \eqs\eqref{eqmom}, the latter can be expressed as a function of $\sigma$ and $\beta$. While $\sigma$ is easily obtained from its equation of motion \eqref{sigmaeom}, the determination of $\beta$ requires solving the equations of motion for the energy-momentum tensor \eqref{leom2} and using \eq\eqref{epsbeta} to obtain $\beta$ from $\varepsilon$. Thus, we  provided a closed set of equations to determine the polarization vector. The explicit numerical calculation is left for future work.

\section{Conclusions}
\label{concsec}

In this paper, we derived equations of motion for the transverse polarization of a boost-invariant system from kinetic theory. Using spherical harmonics to expand the distribution function, we found that the polarization vector can be written as an infinite sum over spin moments $\mG^k_{n\ell r}$ and $\mI^k_{n\ell r}$ with $r=0$. In order to derive equations of motion for the spin moments, we considered the Boltzmann equation for the spin-dependent distribution function with the local collision term modeled by a relaxation time approximation. After imposing boost-invariance in $z$-direction, the left-hand side of the Boltzmann equation contains a term proportional to the $\ms_z$-derivative of the distribution function in addition to the $p_z$-derivative. The local-equilibrium distribution function depends on the spin potential, which has only one nonzero component if we restrict the polarization to the transverse plane. We used a matching condition in order to ensure the microscopic conservation of spin angular momentum, which determines the nonzero component of the spin potential. Then, we derived equations of motion for the spin moments $\mG^k_{n\ell r}$ and $\mI^k_{n\ell r}$. In contrast to the massless case, the free-streaming parts of these equations of motion contain six terms which couple spin moments with different $n$ and with different $r$. We showed that for free streaming, the equations of motion feature an unstable fixed point, where the spin moments become independent of $r$, and a stable fixed point, where the terms which couple spin moments with different $r$ vanish and moments with different $n$ and $\ell$ are proportional to each other. At both free-streaming fixed points, all spin moments decay with power laws. On the other hand, we found that in the hydrodynamic regime only spin moments with $|\ell|=1$ and $n$ odd decay with power laws and therefore feature a power expansion around local equilibrium, while all other spin moments decay exponentially. Using these properties of the equations of motion, we were able to truncate the sum over spin moments in the polarization vector and close the system of equations of motion in terms of the dynamical spin moments. First, since the polarization is measured at freeze out, where exponentially decaying moments have disappeared, we dropped the corresponding terms in the polarization vector. For the remaining terms, we inserted the power expansion of the spin moments around local equilibrium, resulting in an expression for the polarization which is a power series in $w^{-k}$. For any fixed value $k$, the sum over the index $n$ of the spin moments is finite. This allowed us to order the appearing spin moments by the time scales on which they decay in the hydrodynamic regime. Choosing a truncation at a given value of $k$ then also implies a maximal value for $n$ and corresponds to neglecting spin moments which decay at least $\sim w^{-(k+1)}$ faster than the equilibrium spin moments. We then showed how to close the system of moment equations after choosing a truncation by replacing the spin moments with $r=0$ and $n$ larger than the maximal value by its value at the free-streaming point, with this replacement being valid also in local equilibrium. On the other hand, for the spin moments with $r=2$ appearing in the equations of motion for the spin moments with $r=0$ we used an interpolation between the free-streaming fixed point, where the corresponding terms vanish, and the local-equilibrium regime. Finally, we obtained the equation of motion for the equilibrium spin moment $\sigma$ and expressed the other relevant equilibrium spin moments in terms of $\sigma$. As an example, we gave the expression for the polarization using a truncation at $k=1$ and showed the corresponding closed equations of motion for the dynamical spin moments.

Our results show that, while the global polarization for Bjorken symmetry vanishes, the local polarization in the transverse plane can be nonzero. This polarization is not induced by thermal vorticity, which is zero due to the assumption of translational invariance, but emerges from the initial conditions and evolves from the free-streaming regime to the hydrodynamic regime, where it decays with power laws, since, for a local collision term, the dipole moment tensor is a collisional invariant and its components survive on hydrodynamic time scales. Possible sources of an initial polarization of the quark-gluon plasma in heavy-ion collisions are interactions with color fields in the glasma stage, see, e.g., Refs.~\cite{Muller:2021hpe,Kumar:2022ylt}. In the future, it will be interesting to solve the equations of motion derived in this work, and to study in particular the dependence of the result on the initial conditions. One may then compare the results to measurements~\cite{Niida:2018hfw} and local-equilibrium calculations~\cite{Becattini:2021iol} of the momentum dependence of the transverse polarization. Furthermore, it will be an interesting extension of this work to consider longitudinal polarization, as well as a nonlocal collision term. One may also relax the assumption of translational invariance in the $x$-$y$-plane in order to allow for nonzero thermal vorticity.

\section*{Acknowledgements}

N.W.\ acknowledges support by the German National Academy of Sciences Leopoldina through the Leopoldina fellowship program with funding code LPDS 2022-11.

\begin{appendix}

\section{Details for the polarization vector}
\label{coeffapp}

The coefficients in \eq\eqref{pixlong} are given by
\begin{align}    \mathfrak{a}_{n\ell}^+&\equiv\frac14 \frac{1}{2n+1}\left(\frac{1}{2n+3}+\frac{1}{2n-1}\right)\; ,\n\\
    \mathfrak{b}_{n\ell}^+&\equiv-\frac14 \frac{1}{2n+1}\frac{1}{2n-1}\; ,\n\\
    \mathfrak{c}_{n\ell}^+&\equiv-\frac14 \frac{1}{2n+1}\frac{1}{2n+3}\; ,\n\\
    \mathfrak{a}^-_{n\ell}&\equiv -\frac14 \frac{1}{2n+1}(n+\ell)(n+\ell-1)(n-\ell+1)\left(\frac{n-\ell+2}{2n+3}+\frac{n-\ell}{2n-1} \right)\; ,\n\\
    \mathfrak{b}^-_{n\ell}&\equiv\frac14 \frac{1}{2n+1}\frac{1}{2n-1}(n+\ell-1)(n+\ell)(n+\ell-3)(n+\ell-2)\; ,\n\\
    \mathfrak{c}_{n\ell}^-&\equiv \frac14 \frac{1}{2n+1}\frac{1}{2n+3} (n-\ell+1)(n-\ell+2)(n-\ell+3)(n-\ell+4)\; ,\n\\
    \mathfrak{a}^0_{n\ell}&\equiv -\frac12 \frac{1}{2n+1}\left(\frac{(n+2+\ell)(n+1+\ell)}{2n+3}+\frac{(n-\ell-1)(n-\ell)}{2n-1} \right)\; ,\n\\
    \mathfrak{b}^0_{n\ell}&\equiv \frac12 \frac{1}{2n+1} \frac{1}{2n-1}(n+\ell)(n-1+\ell)\; ,\n\\
    \mathfrak{c}^0_{n\ell}&\equiv \frac12 \frac{1}{2n+1} \frac{1}{2n+3} (n-\ell+1)(n-\ell+2)\; ,\n\\
    \mathfrak{d}^+_{n\ell}&\equiv-\frac12\frac{1}{2n+1}\left(\frac{n+\ell+2}{2n+3}-\frac{n-\ell-1}{2n-1}\right)\; ,\n\\
    \mathfrak{e}^+_{n\ell}&\equiv\frac12\frac{1}{2n+1}\frac{n+\ell}{2n-1}\; ,\n\\
    \mathfrak{f}^+_{n\ell}&\equiv -\frac12\frac{1}{2n+1} \frac{n-\ell+1}{2n+3}\; ,\n\\
    \mathfrak{d}^-_{n\ell}&\equiv \frac12\frac{1}{2n+1}\left[(n-\ell+1)(n-\ell+2)\frac{n+\ell}{2n+3}-(n+\ell-1)(n+\ell)\frac{n-\ell+1}{2n-1} \right]\; ,\n\\
    \mathfrak{e}^-_{n\ell}&\equiv -\frac12\frac{1}{2n+1} (n+\ell-1)(n+\ell)\frac{n+\ell-2}{2n-1}\; ,\n\\
    \mathfrak{f}^-_{n\ell}&\equiv \frac12\frac{1}{2n+1} (n-\ell+1)(n-\ell+2)\frac{n-\ell+3}{2n+3}\; .\n\\\label{abcpol}
\end{align}

The calculation in order to obtain \eq\eqref{pixnle} reads
\begin{align}
    \Pi_\star^x(\phi)&= \frac{1}{2\mathcal{N}} \sum_{n=1,3,5,\ldots}\Bigg\{\sum_{\ell=\pm1}\left[N_{n\ell} \mathfrak{I}_{n\ell}\mG^x_{n\ell0}+(N_{n\ell} \mathfrak{I}_{n\ell}\mathfrak{a}^0_{n\ell}+N_{(n+2)\ell} \mathfrak{I}_{(n+2)\ell}\mathfrak{b}_{(n+2)\ell}^0+N_{(n-2)\ell} \mathfrak{I}_{(n-2)\ell}\mathfrak{c}^0_{(n-2)\ell})\right.\n\\
    &\left.\times\left(\mI^x_{n\ell0}-\mG^x_{n\ell0}\right)\right]   e^{i\ell\phi}+\left(\mathfrak{a}^+_{n(-1)}N_{n(-1)} \mathfrak{I}_{n(-1)}+\mathfrak{b}^+_{(n+2)(-1)}N_{(n+2)(-1)} \mathfrak{I}_{(n+2)(-1)}+\mathfrak{c}^+_{(n-2)(-1)}N_{(n-2)(-1)} \mathfrak{I}_{(n-2)(-1)}\right)\n\\
    &\times\left(\mI^x_{n10}-\mG^x_{n10}+\mI^y_{n10}-\mG^y_{n10}\right) e^{-i\phi}\n\\
    &+\left(\mathfrak{a}^-_{n1}N_{n1} \mathfrak{I}_{n1}+\mathfrak{b}^-_{(n+2)1}N_{(n+2)1} \mathfrak{I}_{(n+2)1}+\mathfrak{c}^-_{(n-2)1}N_{(n-2)1} \mathfrak{I}_{(n-2)1}\right)\left(\mI^x_{n(-1)0}-\mG^x_{n(-1)0}+\mI^y_{n(-1)0}-\mG^y_{n(-1)0}\right) e^{i\phi}\Bigg\}\n\\
    &+\frac{1}{2\mathcal{N}} \sum_{n=3,5,\ldots}\bigg[ \left(\mathfrak{a}^+_{n(-3)}N_{n(-3)} \mathfrak{I}_{n(-3)}+\mathfrak{b}^+_{(n+2)(-3)}N_{(n+2)(-3)} \mathfrak{I}_{(n+2)(-3)}+\mathfrak{c}^+_{(n-2)(-3)}N_{(n-2)(-3)} \mathfrak{I}_{(n-2)(-3)}\right)\n\\
    &\times\left(\mI^x_{n(-1)0}-\mG^x_{n(-1)0}+\mI^y_{n(-1)0}-\mG^y_{n(-1)0}\right) e^{-3i\phi}\n\\
    &+\left(\mathfrak{a}^-_{n3}N_{n3} \mathfrak{I}_{n3}+\mathfrak{b}^-_{(n+2)3}N_{(n+2)3} \mathfrak{I}_{(n+2)3}+\mathfrak{c}^-_{(n-2)3}N_{(n-2)3} \mathfrak{I}_{(n-2)3}\right)\left(\mI^x_{n10}-\mG^x_{n10}+\mI^y_{n10}-\mG^y_{n10}\right) e^{3i\phi}\bigg]\n\\
    &=\frac{1}{2\mathcal{N}}  \sum_{k=0}^\infty \sum_{n=1,3,\ldots,2k+1} \left[h_n\mG^x_{110,\text{eq}}  g_{n1}^{x(k)}+\tilde{h}_n\mI^x_{110,\text{eq}} i_{n1}^{x(k)}\right.\n\\
    &\left. -z_n\left(\mG^x_{1(-1)0,\text{eq}}  g_{n(-1)}^{x(k)}-\mI^x_{1(-1)0,\text{eq}} i_{n(-1)}^{x(k)}+\mG^y_{1(-1)0,\text{eq}} g_{n(-1)}^{y(k)}-\mI^y_{1(-1)0,\text{eq}} i_{n(-1)}^{y(k)}\right) \right] w^{-k} e^{i\phi}\n\\
    &+\frac{1}{2\mathcal{N}}  \sum_{k=0}^\infty \sum_{n=1,3,\ldots,2k+1} \left[-\bar{h}_n\left(\mG^x_{110,\text{eq}}  g_{n1}^{x(k)}-\mI^x_{110,\text{eq}} i_{n1}^{x(k)}+\mG^y_{110,\text{eq}} g_{n1}^{y(k)}-\mI^y_{110,\text{eq}} i_{n1}^{y(k)}\right)\right.\n\\
    & \left. +\bar{z}_n\mG^x_{1(-1)0,\text{eq}}  g_{n(-1)}^{x(k)}+\hat{z}_n\mI^x_{1(-1)0,\text{eq}} i_{n(-1)}^{x(k)}\right] w^{-k} e^{-i\phi}\n\\
    &-\frac{1}{2\mathcal{N}}  \sum_{k=1}^\infty \sum_{n=1,3,\ldots,2k+1} \hat{q}_n\left(\mG^x_{110,\text{eq}}  g_{n1}^{x(k)}-\mI^x_{110,\text{eq}} i_{n1}^{x(k)}+\mG^y_{110,\text{eq}} g_{n1}^{y(k)}-\mI^y_{110,\text{eq}} i_{n1}^{y(k)}\right)w^{-k} e^{3i\phi}\n\\
    &-\frac{1}{2\mathcal{N}}  \sum_{k=1}^\infty \sum_{n=3,5,\ldots,2k+1} \hat{v}_n \left(\mG^x_{1(-1)0,\text{eq}}  g_{n(-1)}^{x(k)}-\mI^x_{1(-1)0,\text{eq}} i_{n(-1)}^{x(k)}+\mG^y_{1(-1)0,\text{eq}} g_{n(-1)}^{y(k)}-\mI^y_{1(-1)0,\text{eq}} i_{n(-1)}^{y(k)}\right) w^{-k} e^{-3i\phi}\n\\
    &= \frac{1}{\mathcal{N}} \, \text{Re}\, \Bigg\{ \sum_{k=0}^\infty \sum_{n=1,3,\ldots,2k+1} \left[\left(h_n\mG^x_{110,\text{eq}}  g_{n1}^{x(k)}+\tilde{h}_n\mI^x_{110,\text{eq}} i_{n1}^{x(k)}\right)e^{i\phi}\right.\n\\
    &\left. -\bar{h}_n\left(\mG^x_{110,\text{eq}}  g_{n1}^{x(k)}-\mI^x_{110,\text{eq}} i_{n1}^{x(k)}+\mG^y_{110,\text{eq}} g_{n1}^{y(k)}-\mI^y_{110,\text{eq}} i_{n1}^{y(k)}\right) e^{-i\phi} \right] w^{-k} \Bigg\}\n\\
    &-\frac{1}{\mathcal{N}}  \, \text{Re}\, \Bigg[\sum_{k=1}^\infty \sum_{n=3,5,\ldots,2k+1} \hat{q}_n\left(\mG^x_{110,\text{eq}}  g_{n1}^{x(k)}-\mI^x_{110,\text{eq}} i_{n1}^{x(k)}+\mG^y_{110,\text{eq}} g_{n1}^{y(k)}-\mI^y_{110,\text{eq}} i_{n1}^{y(k)}\right)w^{-k} e^{3i\phi}\Bigg] 
\end{align}
with $\mathfrak{a}^\pm_{n\ell}\equiv0$ for $|\ell|>n$, similar for all other coefficients. We also defined
\begin{equation}
    \mathfrak{I}_{n\ell}\equiv \int d\cos\theta\, \Pc_n^\ell(\cos\theta)\; . \label{iint}
\end{equation}
The coefficients are obtained as
\begin{align}
    h_n&\equiv N_{n1} \mathfrak{I}_{n1} (1-\mathfrak{a}^0_{n1})-N_{(n+2)1} \mathfrak{I}_{(n+2)1}\mathfrak{b}_{(n+2)1}^0-N_{(n-2)1} \mathfrak{I}_{(n-2)1}\mathfrak{c}^0_{(n-2)1}\; ,\n\\
    \tilde{h}_n&\equiv N_{n1} \mathfrak{I}_{n1} \mathfrak{a}^0_{n1}+N_{(n+2)1} \mathfrak{I}_{(n+2)1}\mathfrak{b}_{(n+2)1}^0+N_{(n-2)1} \mathfrak{I}_{(n-2)1}\mathfrak{c}^0_{(n-2)1}\; ,\n\\
    %q_n&\equiv 0\; ,\n\\
    z_n&\equiv  \mathfrak{a}^-_{n1}N_{n1} \mathfrak{I}_{n1}+\mathfrak{b}^-_{(n+2)1}N_{(n+2)1} \mathfrak{I}_{(n+2)1}+\mathfrak{c}^-_{(n-2)1}N_{(n-2)1} \mathfrak{I}_{(n-2)1}\n\\
    \bar{h}_n&\equiv \mathfrak{a}^+_{n(-1)}N_{n(-1)} \mathfrak{I}_{n(-1)}+\mathfrak{b}^+_{(n+2)(-1)}N_{(n+2)(-1)} \mathfrak{I}_{(n+2)(-1)}+\mathfrak{c}^+_{(n-2)(-1)}N_{(n-2)(-1)} \mathfrak{I}_{(n-2)(-1)}\; ,\n\\
    \bar{z}_n&\equiv N_{n(-1)} \mathfrak{I}_{n(-1)} (1-\mathfrak{a}^0_{n(-1)})-N_{(n+2)(-1)} \mathfrak{I}_{(n+2)(-1)}\mathfrak{b}_{(n+2)(-1)}^0-N_{(n-2)(-1)} \mathfrak{I}_{(n-2)(-1)}\mathfrak{c}^0_{(n-2)(-1)}\; ,\n\\
    \hat{z}_n&\equiv N_{n(-1)} \mathfrak{I}_{n(-1)} \mathfrak{a}^0_{n(-1)}+N_{(n+2)(-1)} \mathfrak{I}_{(n+2)(-1)}\mathfrak{b}_{(n+2)(-1)}^0+N_{(n-2)(-1)} \mathfrak{I}_{(n-2)(-1)}\mathfrak{c}^0_{(n-2)(-1)}\; ,\n\\
    %\bar{v}_n&= 0\; ,\n\\
    \hat{q}_n&\equiv \mathfrak{a}^-_{n3}N_{n3} \mathfrak{I}_{n3}+\mathfrak{b}^-_{(n+2)3}N_{(n+2)3} \mathfrak{I}_{(n+2)3}+\mathfrak{c}^-_{(n-2)3}N_{(n-2)3} \mathfrak{I}_{(n-2)3}\; ,\n\\
    \hat{v}_n&\equiv \mathfrak{a}^+_{n(-3)}N_{n(-3)} \mathfrak{I}_{n(-3)}+\mathfrak{b}^+_{(n+2)(-3)}N_{(n+2)(-3)} \mathfrak{I}_{(n+2)(-3)}+\mathfrak{c}^+_{(n-2)(-3)}N_{(n-2)(-3)} \mathfrak{I}_{(n-2)(-3)}\; .
    \label{coeffhz}
\end{align}

\section{Exact solution of the Boltzmann equation for free streaming}
\label{freeapp}

Consider \eq\eqref{boltzbjorkcomp} for free streaming, i.e.,
\begin{equation}
    \left(\partial_\tau-\frac{p_z}{\tau}\partial_{p_z}-\frac{p_z}{E_p^2}\frac{\mathbf{p}\cdot \boldsymbol{\ms}}{\tau}\partial_{\ms_z}\right)f(\tau,\p_\bot,p_z,\theta_\ms,\ms_z)=0 .\label{freeboltz}
\end{equation}
Using the initial condition at $\tau=\tau_0$
\begin{equation}
   f(\tau=\tau_0)= f_\text{in}(\p_\bot,p_z,\ms_x,\ms_y),
\end{equation}
we find the following solution
\begin{equation}
f(\tau,\p_\bot,p_z,\theta_\ms,\ms_z)= f_\text{in}\left(\p_\bot,p_z\frac{\tau}{\tau_0},\msb\cos\theta_\ms,\msb\sin\theta_\ms\right) \; , \label{freesol0}
\end{equation}
i.e., the free streaming formally does not affect the polarization. However, since $\msb$ depends on $p_z$, the change in the momentum distribution has an implicit effect on the polarization.
In order to prove that \eq\eqref{freesol0} is a solution of \eq\eqref{freeboltz}, we note that the derivative of $f_\text{in}$ with respect to its second argument will be multiplied by
\begin{equation}
    \left(\partial_\tau-\frac{p_z}{\tau}\partial_{p_z}\right)p_z\frac{\tau}{\tau_0}=0\; .
\end{equation}
Furthermore, using
\begin{align}
    \partial_{p_z} \msb&=\frac{1}{\msb}\frac{\boldsymbol{\ms}\cdot\mathbf{p}}{E_p}\left(\frac{\ms_z}{E_p}-\frac{\boldsymbol{\ms}\cdot\mathbf{p}}{E_p^3}p_z\right)+\frac{1}{\msb}\frac{\boldsymbol{\ms}\cdot\mathbf{p}}{E_p^2}(p_x\cos\theta_\ms+p_y\sin\theta_\ms)\partial_{p_z}\msb\; , \n\\ \partial_{\ms_z} \msb &=\frac{1}{\msb} \left(\frac{\boldsymbol{\ms}\cdot\mathbf{p}}{E_p^2}p_z-\ms_z \right)+\frac{1}{\msb}\frac{\boldsymbol{\ms}\cdot\mathbf{p}}{E_p^2}(p_x\cos\theta_\ms+p_y\sin\theta_\ms)\partial_{\ms_z}\msb\; ,
\end{align}
we find that the derivative of $f_\text{in}$ with respect to its third or fourth argument, respectively, is multiplied by
\begin{align}
\left(-\frac{p_z}{\tau}\partial_{p_z}-\frac{p_z}{E_p^2}\frac{\mathbf{p}\cdot \boldsymbol{\ms}}{\tau}\partial_{\ms_z}\right)\msb = & \left[1-\frac{1}{\msb}\frac{\boldsymbol{\ms}\cdot\mathbf{p}}{E_p^2}(p_x\cos\theta_\ms+p_y\sin\theta_\ms)\right]^{-1}\n\\
&\times\left(-\frac{p_z}{\tau}\frac{\ms_z}{\msb} \frac{\boldsymbol{\ms}\cdot\mathbf{p}}{E_p^2}+\frac{p_z^2}{\tau}\frac{1}{\msb} \frac{(\boldsymbol{\ms}\cdot\mathbf{p})^2}{E_p^4}-\frac{p_z^2}{\tau} \frac{1}{\msb} \frac{(\boldsymbol{\ms}\cdot\mathbf{p})^2}{E_p^4}+\frac{p_z}{\tau} \frac{\boldsymbol{\ms}\cdot\mathbf{p}}{E_p^2} \frac{\ms_z}{\tau}\right)\n\\
=&0\; . \label{dsx0}
\end{align}

We can use the exact free-streaming solution in order to analyze the free-streaming fixed points of the spin moments. Consider
\begin{align}
    \mG^x_{n\ell r}&= \int_{p\ms} \msb \sin\theta_\ms \left(\frac{p}{E_p}\right)^r  \Pc_n^\ell(p_z/p) e^{i\ell\phi} f_\text{in}\left(p_\bot,p_z\frac{\tau}{\tau_0},\msb\cos\theta_\ms,\msb\sin\theta_\ms\right)\n\\
    &= 2\int_{p} E_p \left(\frac{p}{E_p}\right)^r \Pc_n^\ell(p_z/p) e^{i\ell\phi}\left[\left(1+\frac{p_x^2}{m^2}\right) \A_\text{in}^x\left(\p_\bot, p_z\frac{\tau}{\tau_0}\right)+\frac{p_x p_y}{m^2} \A_\text{in}^y\left(\p_\bot, p_z\frac{\tau}{\tau_0}\right)\right]\n\\
     &= 2\frac{\tau_0}{\tau}\int_{p} \epsilon_{p\tau}\left(\frac{p_\tau}{\epsilon_{p\tau}}\right)^r \Pc_n^\ell(\tau_0p_z/\tau p_\tau) e^{i\ell\phi}\left[\left(1+\frac{p_x^2}{m^2}\right) \A_\text{in}^x\left(\p_\bot, p_z\right)+\frac{p_x p_y}{m^2} \A_\text{in}^y\left(\p_\bot, p_z\right)\right]\; .
    \end{align}
where we inserted \eq\eqref{ftransvpol}, used $\int dS(\p)=2$, $\int dS(\p)\, \ms^i=0$, and $\int dS(\p)\, \ms^i \ms^j=2(\delta^{ij}+p^i p^j/m^2)$, and defined $\int_p \equiv \int d^3p$, $\epsilon_{p\tau}\equiv \sqrt{(\tau_0 p_z/\tau)^2+\p_\bot^2+m^2}$ and $p_{\tau}\equiv \sqrt{(\tau_0 p_z/\tau)^2+\p_\bot^2}$. For early time $\tau\ll\tau_0$ we find
\begin{align}
    \mG^x_{n\ell r} &\rightarrow 2\left(\frac{\tau_0}{\tau}\right)^{2-\ell}\int_{p} |p_z| [\text{sgn}(p_z)]^{n+\ell}   \left( \frac{p_\bot}{|p_z|}\right)^\ell B^\ell_n e^{i\ell\phi} \left[\left(1+\frac{p_x^2}{m^2}\right) \A_\text{in}^x\left(\p_\bot, p_z\right)+\frac{p_x p_y}{m^2} \A_\text{in}^y\left(\p_\bot, p_z\right)\right]\n\\
    &\sim \left(\frac{1}{\tau}\right)^{2-\ell}\; 
\end{align} 
with
\begin{equation}
    B_{n\ell}\equiv \lim_{x\rightarrow 1} \frac{\Pc_n^\ell(x)}{{\sqrt{1-x^2}}^\ell}\; .
    \label{bnell}
\end{equation}
The dependence on $r$ vanishes in this limit.
For late time $\tau\gg\tau_0$ we obtain
\begin{align}
    \mG^x_{n\ell r} &\rightarrow 2\frac{\tau_0}{\tau}\int_{p} \varepsilon_{\bot} \left(\frac{p_\bot}{\varepsilon_\bot}\right)^r \mathcal{P}^\ell_n(0)  e^{i\ell\phi}\left[\left(1+\frac{p_x^2}{m^2}\right) \A_\text{in}^x\left(\p_\bot, p_z\right)+\frac{p_x p_y}{m^2} \A_\text{in}^y\left(\p_\bot, p_z\right)\right]\sim \frac{1}{\tau}\; , \qquad \qquad \qquad n+\ell \text{ even}\n\\
        \mG^x_{n\ell r} &\rightarrow 2\left(\frac{\tau_0}{\tau}\right)^2\int_{p} p_z \left(\frac{p_\bot}{\varepsilon_\bot}\right)^{r-1}  (\mathcal{P}^\ell_n)^\prime(0)  e^{i\ell\phi} \left[\left(1+\frac{p_x^2}{m^2}\right) \A_\text{in}^x\left(\p_\bot, p_z\right)+\frac{p_x p_y}{m^2} \A_\text{in}^y\left(\p_\bot, p_z\right)\right]\sim \left(\frac{1}{\tau}\right)^2\; ,  \; n+\ell \text{ odd} \; .
\end{align}
Here we defined $\varepsilon_\bot\equiv \sqrt{p_\bot^2+m^2}$ and used the recurrence relations
\begin{align}
    \Pc_n^\ell(0)&=-\frac{n-1+\ell}{n-\ell}\Pc_{n-2}^\ell(0)\; , & n>1,\; \ell\neq n\n\\
    \Pc_n^\ell(0)&=-(n+\ell-1)(n-\ell+2) \Pc_n^{\ell-2}(0)\; , & \ell>1
\end{align}
and
\begin{align}
    \Pc_0^0(0)&=1\; ,& \Pc_1^0(0)&= 0\; ,& \Pc_1^1(0)&=-1\; ,& \Pc_0^1(0)&=0
\end{align}
to find that $\Pc_n^\ell(0)=0$ for $n+\ell$ odd. We also have
\begin{equation}
    (\Pc_n^\ell)^\prime(0)=(n+\ell)\Pc_{n-1}^\ell(0)\; .
\end{equation}

The behavior of $\mI^x_{n\ell r}$ is obtained analogously as
\begin{align}
    \mI^x_{n\ell r} &\rightarrow 2m\left(\frac{\tau_0}{\tau}\right)^{1-\ell}\int_{p} \, [\text{sgn}(p_z)]^{n+\ell}   \left( \frac{p_\bot}{|p_z|}\right)^\ell B^\ell_n e^{i\ell\phi} \left[\left(1+\frac{p_x^2}{m^2}\right) \A_\text{in}^x\left(\p_\bot, p_z\right)+\frac{p_x p_y}{m^2} \A_\text{in}^y\left(\p_\bot, p_z\right)\right]\n\\
    &\sim \left(\frac{1}{\tau}\right)^{1-\ell}\;
\end{align} 
for $\tau\ll\tau_0$ and 
\begin{align}
    \mI^x_{n\ell r} &\rightarrow 2m\frac{\tau_0}{\tau}\int_{p}  \left(\frac{p_\bot}{\varepsilon_\bot}\right)^r \mathcal{P}^\ell_n(0)  e^{i\ell\phi}\left[\left(1+\frac{p_x^2}{m^2}\right) \A_\text{in}^x\left(\p_\bot, p_z\right)+\frac{p_x p_y}{m^2} \A_\text{in}^y\left(\p_\bot, p_z\right)\right]\sim \frac{1}{\tau}\; , \qquad \qquad \qquad n+\ell \text{ even}\n\\
        \mI^x_{n\ell r} &\rightarrow 2m\left(\frac{\tau_0}{\tau}\right)^2\int_{p} \frac{p_z}{\varepsilon_\bot} \left(\frac{p_\bot}{\varepsilon_\bot}\right)^{r-1}  (\mathcal{P}^\ell_n)^\prime(0)  e^{i\ell\phi} \left[\left(1+\frac{p_x^2}{m^2}\right) \A_\text{in}^x\left(\p_\bot, p_z\right)+\frac{p_x p_y}{m^2} \A_\text{in}^y\left(\p_\bot, p_z\right)\right]\sim \left(\frac{1}{\tau}\right)^2\; ,  \; n+\ell \text{ odd} \; .
\end{align}
for $\tau\gg\tau_0$.

\section{Free-streaming fixed points of the equations of motion}
\label{ffpapp}

For $\tau\ll\tau_R$, the equations of motion \eqref{geom} and \eqref{ieom} are dominated by the terms in the first lines, respectively, corresponding to free streaming. The equations of motion feature two free-streaming fixed points. The unstable fixed point is obtained from the relation
\begin{align}
    B_{n\ell} (a_{n\ell r}+d_{n\ell r})+B_{(n-2)\ell} (b_{n\ell r}+e_{n\ell r})+B_{(n+2)\ell}(c_{n\ell r}+f_{n\ell r})&= 2-\ell\; , \label{anotherrelation}
\end{align}
where $B_{n\ell}$ is defined in \eq\eqref{bnell}.
At this fixed point, the spin moments become independent of $r$ and related via $\mG^x_{n\ell r}/\mG^x_{m k s}=B_{n\ell}/B_{m k}$ for $m\geq k$. They decay $\sim \tau^{-2+\ell}$. Furthermore, the identities
\begin{align}
    {a}_{n\ell r}  \mathcal{P}^\ell_{n}(0)+{b}_{n\ell r}\mathcal{P}^\ell_{n-2}(0)+{c}_{n\ell r}\mathcal{P}^\ell_{n+2}(0)&=\mathcal{P}^\ell_{n}(0)\; , & n+\ell \text{ even},\n\\
     {d}_{n\ell r}\mathcal{P}^\ell_{n}(0)+{e}_{n\ell r}\mathcal{P}^\ell_{n-2}(0)+{f}_{n\ell r}\mathcal{P}_{n+2}^\ell(0)&=0 & n+\ell \text{ even}\; ,\n\\
    {a}_{n\ell r} (n+\ell) \mathcal{P}^\ell_{n-1}(0)+{b}_{n\ell r}(n-2+\ell)\mathcal{P}^\ell_{n-3}(0)+{c}_{n\ell r}(n+2+\ell)\mathcal{P}^\ell_{n+1}(0)&=2(n+\ell)\mathcal{P}^\ell_{n-1}(0)\; , & n+\ell \text{ odd} \; ,
    \n\\
    {d}_{n\ell r} (n+\ell) \mathcal{P}^\ell_{n-1}(0)+{e}_{n\ell r}(n-2+\ell)\mathcal{P}^\ell_{n-3}(0)+{f}_{n\ell r}(n+2+\ell)\mathcal{P}^\ell_{n+1}(0)&=0\; , & n+\ell \text{ odd}\; .
    \label{stabfprel}
\end{align}
are related to a stable fixed point, where
the equations of motion for moments with different $r$ decouple. The moments are related through
\begin{align}
   \mG^x_{n\ell r}/\mG^x_{mkr}&=\Pc_n^\ell(0)/\Pc_m^k(0)\; , & n+\ell \text{ even}\; ,\n\\
   \mG^x_{n\ell r}/\mG^x_{mkr}&=[(n+\ell)\Pc_{n-1}^\ell(0)/(m+k)\Pc_{m-1}^k(0)]\; , & n+\ell \text{ odd} \; .
   \label{fsfpapp}
\end{align}
The behavior at the free-streaming fixed points can also be analyzed by considering the exact free-streaming solution. This is shown in Appendix \ref{freeapp}.

 For the $\mI$-moments, the identity for the unstable fixed point reads
\begin{align}
    B_{n\ell} (a_{n\ell r}+\tilde{d}_{n\ell r})+B_{(n-2)\ell} (b_{n\ell r}+\tilde{e}_{n\ell r})+B_{(n+2)\ell}(c_{n\ell r}+\tilde{f}_{n\ell r})&= 1-\ell\; , \label{anotherrelation2}
\end{align}
therefore, these moments decay at the unstable free-streaming fixed point $\sim \tau^{-1+\ell}$ with $\mI_{n\ell r}/\mI_{m k s}=B_{n\ell}/B_{m k}$ for $m\geq k$. Furthermore, we also have
\begin{align}
    \tilde{d}_{n\ell r}\mathcal{P}^\ell_{n}(0)+\tilde{e}_{n\ell r}\mathcal{P}^\ell_{n-2}(0)+\tilde{f}_{n\ell r}\mathcal{P}_{n+2}^\ell(0)&=0 & n+\ell \text{ even}\; ,\n\\
    \tilde{d}_{n\ell r} (n+\ell) \mathcal{P}^\ell_{n-1}(0)+\tilde{e}_{n\ell r}(n-2+\ell)\mathcal{P}^\ell_{n-3}(0)+\tilde{f}_{n\ell r}(n+2+\ell)\mathcal{P}^\ell_{n+1}(0)&=0\; , & n+\ell \text{ odd}\; .
\end{align}
Thus, the $\mI$-moments behave identically to the $\mG$-moments at the stable free-streaming fixed point.

\section{Equations of motion for the spin moments}
\label{peomapp}

In order to derive \eq\eqref{geom} from \eq\eqref{boltzbjorkcomp} we use the following calculation for the free-streaming part,
\begin{align}
    \partial_\tau \mG^x_{n\ell r}=&  \int_{p\ms} \left(\frac{p}{E_p}\right)^r \ms^x \mathcal{P}_n^\ell(\cos\theta) e^{i\ell\phi} \left(\frac{p_z}{\tau}\partial_{p_z}+\frac{p_z}{E_p^2}\frac{\mathbf{p}\cdot \boldsymbol{\ms}}{\tau}\partial_{\ms_z}\right)f\n\\
    =& \frac12\int_{p\ms} \left(\frac{p}{E_p}\right)^r \ms^x \mathcal{P}_n^\ell(\cos\theta) e^{i\ell\phi} \left(\frac{p_z}{\tau}\partial_{p_z}+\frac{p_z}{E_p^2}\frac{\mathbf{p}\cdot \boldsymbol{\ms}}{\tau}\partial_{\ms_z}\right)(\F+\ms^x \A^x+\ms^y \A^y)\n\\
    =& \frac12\int_{p\ms} \left(\frac{p}{E_p}\right)^r \ms^x \mathcal{P}_n^\ell(\cos\theta) e^{i\ell\phi} \left(\ms^x\frac{p_z}{\tau}\partial_{p_z} \A^x+\ms^y\frac{p_z}{\tau}\partial_{p_z} \A^y\right)\n\\
    =& 2 \int_{p} E_p \left(\frac{p}{E_p}\right)^r  \mathcal{P}_n^\ell(\cos\theta) e^{i\ell\phi} \left[\left(1+\frac{p_x^2}{m^2}\right)\frac{p_z}{\tau}\partial_{p_z} \A^x+\frac{p_x p_y}{m^2}\frac{p_z}{\tau}\partial_{p_z} \A^y\right]\n\\
    =& -2\frac{1}{\tau} \int_{p} E_p \left(\frac{p}{E_p}\right)^r \left[1-(r-1)\frac{p_z^2}{E_p^2}+r\frac{p_z^2}{p^2}\right]  \mathcal{P}_n^\ell(\cos\theta) e^{i\ell\phi} \left[\left(1+\frac{p_x^2}{m^2}\right) \A^x+\frac{p_x p_y}{m^2} \A^y\right]\n\\
    &-2\frac1\tau \int_{p} E_p \left(\frac{p}{E_p}\right)^r  \cos\theta \left(1-\cos^2\theta\right) (\mathcal{P}_n^\ell)^\prime(\cos\theta) e^{i\ell\phi} \left[\left(1+\frac{p_x^2}{m^2}\right) \A^x+\frac{p_x p_y}{m^2} \A^y\right]\n\\
    =& -\frac{1}{\tau} \int_{p\ms} \left(\frac{p}{E_p}\right)^r \ms^x \mathcal{P}_n^\ell(\cos\theta) e^{i\ell\phi} f-\frac{1}{\tau}\int_{p\ms} \left(\frac{p}{E_p}\right)^r \left[r-(r-1)\frac{p^2}{E_p^2} \right] \ms^x \cos^2\theta\,  \mathcal{P}_n^\ell(\cos\theta) e^{i\ell\phi}  f\n\\ &+\frac{1}{\tau}\int_{p\ms}  \left(\frac{p}{E_p}\right)^r \ms^x  \cos\theta\left[ n\cos\theta\mathcal{P}^\ell_n(\cos\theta)-(n+\ell)\mathcal{P}_{(n-1)}^\ell(\cos\theta)\right]  e^{i\ell\phi} f\n\\
    =& -\frac{1}{\tau} \left( {a}_{n\ell r} \mG^x_{n\ell r}+ {b}_{n\ell r} \mG^x_{(n-2)\ell r}+{c}_{n\ell r} \mG^x_{(n+2)\ell r}+{d}_{n\ell r} \mG^x_{n\ell(r+2)}+{e}_{n\ell r} \mG^x_{(n-2)\ell(r+2)}+{f}_{n\ell r} \mG^x_{(n+2)\ell(r+2)} \right)\; ,
\end{align}
where we inserted \eq\eqref{ftransvpol} and used $\int dS(\p)=2$, $\int dS(\p)\, \ms^i=0$, and $\int dS(\p)\, \ms^i \ms^j=2(\delta^{ij}+p^i p^j/m^2)$.
The coefficients in read
\begin{align}
 {a}_{n\ell r} &\equiv 1-(n-r)\left(\frac{(n+1-\ell)(n+1+\ell)}{(2n+1)(2n+3)}+\frac{(n+\ell)(n-\ell)}{(2n+1)(2n-1)}\right)+\frac{(n+\ell)(n-\ell)}{2n-1}\; ,\n\\
 {b}_{n\ell r} & \equiv -(n-r) \frac{(n+\ell)(n-1+\ell)}{(2n+1)(2n-1)}+\frac{(n+\ell)(n-1+\ell)}{2n-1}\; ,\n\\
 {c}_{n\ell r} &\equiv -(n-r) \frac{(n+1-\ell)(n+2-\ell)}{(2n+1)(2n+3)}\; ,\n\\
  {d}_{n\ell r} &\equiv -(r-1)\left(\frac{(n+1-\ell)(n+1+\ell)}{(2n+1)(2n+3)}+\frac{(n+\ell)(n-\ell)}{(2n+1)(2n-1)}\right)\; ,\n\\
 {e}_{n\ell r} & \equiv -(r-1) \frac{(n+\ell)(n-1+\ell)}{(2n+1)(2n-1)}\; ,\n\\
 {f}_{n\ell r} &\equiv -(r-1) \frac{(n+1-\ell)(n+2-\ell)}{(2n+1)(2n+3)}\; .\label{abcdefbar}
\end{align}
Furthermore, the free-streaming part of \eq\eqref{ieom} is obtained as
\begin{align}
    \partial_\tau \mI^x_{n\ell r}=& m\int_{p\ms} \frac{1}{E_p} \left(\frac{p}{E_p}\right)^r \ms^x \mathcal{P}^\ell_n(\cos\theta) e^{i\ell\phi} \left(\frac{p_z}{\tau}\partial_{p_z}+\frac{p_z}{E_p^2}\frac{\mathbf{p}\cdot \boldsymbol{\ms}}{\tau}\partial_{\ms_z}\right)f\n\\
    =&\frac12 m \int_{p\ms} \frac{1}{E_p} \left(\frac{p}{E_p}\right)^r \ms^x \mathcal{P}^\ell_n(\cos\theta) e^{i\ell\phi} \left(\frac{p_z}{\tau}\partial_{p_z}+\frac{p_z}{E_p^2}\frac{\mathbf{p}\cdot \boldsymbol{\ms}}{\tau}\partial_{\ms_z}\right)(\F+\ms^x \A^x+\ms^y \A^y)\n\\
    =& \frac12m\int_{p\ms} \frac{1}{E_p} \left(\frac{p}{E_p}\right)^r \ms^x \mathcal{P}^\ell_n(\cos\theta) e^{i\ell\phi} \frac{p_z}{\tau}\left(\ms^x\partial_{p_z} \A^x+\ms^y\partial_{p_z}\A^y \right)\n\\
    =& 2m \int_{p}  \left(\frac{p}{E_p}\right)^r  \mathcal{P}^\ell_n(\cos\theta) e^{i\ell\phi} \frac{p_z}{\tau} \left[\left(1+\frac{p_x^2}{m^2}\right)\partial_{p_z} \A^x+\frac{p_x p_y}{m^2}\partial_{p_z} \A^y\right]\n\\
    =& -2m\frac{1}{\tau} \int_{p}  \left(\frac{p}{E_p}\right)^r  \mathcal{P}^\ell_n(\cos\theta) e^{i\ell\phi} \left[1-r\frac{p_z^2}{E_p^2}+r\frac{p_z^2}{p^2}\right]  \left[\left(1+\frac{p_x^2}{m^2}\right) \A^x+\frac{p_x p_y}{m^2} \A^y\right]\n\\
    &-2m\frac{1}{\tau}\int_{p}  \left(\frac{p}{E_p}\right)^r  (\mathcal{P}^\ell_n)^\prime(\cos\theta) e^{i\ell\phi}  \cos\theta \left(1-\cos^2\theta\right)  \left[\left(1+\frac{p_x^2}{m^2}\right) \A^x+\frac{p_x p_y}{m^2} \A^y\right]\n\\
    =& -\frac{1}{\tau}m\int_{p\ms} \frac{1}{E_p}\, \ms^x  \left(\frac{p}{E_p}\right)^r  \mathcal{P}^\ell_n(\cos\theta) e^{i\ell\phi} f-\frac{1}{\tau}m\int_{p\ms} \frac{1}{E_p}\,  \ms^x \left(\frac{p}{E_p}\right)^r  \mathcal{P}^\ell_n(\cos\theta) e^{i\ell\phi} \cos^2\theta \left[r-r\frac{p^2}{E_p^2}\right] f\n\\
    &+\frac{1}{\tau}m\int_{p\ms} \frac{1}{E_p}\,  \ms^x \left(\frac{p}{E_p}\right)^r  e^{i\ell\phi}\cos\theta\left[ n\cos\theta\, \mathcal{P}_n^\ell(\cos\theta)-(n+\ell)\mathcal{P}^\ell_{(n-1)}(\cos\theta)\right]  f\n\\
    &= -\frac{1}{\tau} \left( {a}_{n\ell r} \mI^x_{n\ell r}+ {b}_{n\ell r} \mI^x_{(n-2)\ell r}+{c}_{n\ell r} \mI^x_{(n+2)\ell r}+\tilde{d}_{n\ell r} \mI^x_{n\ell(r+2)}+\tilde{e}_{n\ell r} \mI^x_{(n-2)\ell(r+2)}+\tilde{f}_{n\ell r} \mI^x_{(n+2)\ell(r+2)} \right)\; . 
        \label{eomixgenpol}
\end{align}
We see that the only difference compared to $\mG^x_{n\ell r}$ appears in the last three terms. The coefficients are given by
\begin{align}
 \tilde{d}_{n\ell r} &\equiv -r\left(\frac{(n+1-\ell)(n+1+\ell)}{(2n+1)(2n+3)}+\frac{(n+\ell)(n-\ell)}{(2n+1)(2n-1)}\right)\; ,\n\\
 \tilde{e}_{n\ell r} & \equiv -r \frac{(n+\ell)(n-1+\ell)}{(2n+1)(2n-1)}\; ,\n\\
 \tilde{f}_{n\ell r} &\equiv -r \frac{(n+1-\ell)(n+2-\ell)}{(2n+1)(2n+3)} \;.   
\end{align}

\section{Equations of motion for the energy-momentum tensor}
\label{emtensapp}

In order to determine the temperature, we need to consider the equation of motion for the component $T^{00}$ of the energy-momentum tensor, cf.\ Ref.~\cite{Jaiswal:2022udf}. We define the moments
\begin{equation}
    \mathcal{L}_{nr}\equiv \frac14\int_{p\ms} \left(\frac{p}{E_p}\right)^r \Pc_n(\cos\theta) f(\tau,\p,\ms) = \int d^3p\, \left(\frac{p}{E_p}\right)^r E_p\, \Pc_n(\cos\theta) \F(\tau,\p)\; .
\end{equation}
We assume that $\F$ is invariant under parity and depends on $p_x$ and $p_y$ only through $p_\bot\equiv \sqrt{p_x^2+p_y^2}$, thus we do not need to take into account any dependence on $\phi$ and can consider moments defined only in terms of Legendre polynomials $\Pc_n(\cos\theta)\equiv \Pc_n^0(\cos\theta)$. The independent components of the energy-momentum tensor are
\begin{align}
 \varepsilon&\equiv\int d^3p\, E_p \F= \mL_{00}\; ,\n\\
 P_L&\equiv \int d^3p\, \frac{p_z^2}{E_p} \F= \frac13 \left( 2\mL_{22}+\mL_{02} \right)\; ,\n\\
 P_T&\equiv \int d^2p\, \frac{p_\bot^2}{E_p}\F=\frac23 \left( \mL_{02}-\mL_{22} \right)\; ,
\end{align}
where $\varepsilon$ is the energy density, $P_L$ is the longitudinal pressure and $P_T$ is the transverse pressure.
 The equations of motion for the moments $\mL_{nr}$ read
\begin{align}
    \partial_\tau \mL_{nr}
    =& -\frac{1}{\tau} \left( {a}_{n0 r} \mL_{n r}+ {b}_{n0 r} \mL_{(n-2)r}+{c}_{n0 r} \mL_{(n+2) r}+{d}_{n0 r} \mL_{n(r+2)}+{e}_{n0 r} \mL_{(n-2)(r+2)}+{f}_{n0 r} \mL_{(n+2)(r+2)} \right)\n\\
    &-\frac{1}{\tau_R} \left(\mL_{n r}-\mL_{n r,\text{eq}}\right)\; ,
    \label{leom}
\end{align}
where the coefficients are equal to those in \eq\eqref{geom} for $\ell=0$. Due to the matching condition \eqref{matching}, we have $\mL_{00}=\mL_{00,\text{eq}}$. Furthermore $\mL_{22,\text{eq}}=0$. We are then left with the equations of motion
\begin{align}
\partial_\tau \mL_{00}
    =& -\frac{1}{\tau} \left( {a}_{000} \mL_{00}+{d}_{000} \mL_{02}+{f}_{000} \mL_{22} \right)\; ,\n\\
    \partial_\tau \mL_{02}
    =& -\frac{1}{\tau} \left( {a}_{002} \mL_{02}+{c}_{002} \mL_{22}+{d}_{002} \mL_{04}+{f}_{002} \mL_{24} \right)-\frac{1}{\tau_R} \left(\mL_{02}-\mL_{02,\text{eq}}\right)  \; ,\n\\
    \partial_\tau \mL_{22}
    =& -\frac{1}{\tau} \left( {a}_{202} \mL_{22}+b_{202}\mL_{02}+{d}_{202} \mL_{24}+e_{202}\mL_{04}+{f}_{202} \mL_{44} \right)-\frac{1}{\tau_R} \mL_{22} \; .
\end{align}
Note that $c_{000}={c}_{202}=0$. Furthermore $a_{000}=1$, $d_{000}=1/3$, and $f_{000}=2/3$. In order to close the system of equations, we need approximations for $\mL_{04}$, $\mL_{24}$, and $\mL_{44}$. We note that these moments appear only in terms which vanish at the free-streaming fixed point. Therefore, we need to take into account only the equilibrium contribution, which is nonzero only for $\mL_{04}$. Again using an interpolation analogous to Ref.~\cite{Jaiswal:2022udf}, we obtain
\begin{align}
\partial_w \mL_{00}
    =& -\frac{1}{w} \left(  \mL_{00}+\frac13 \mL_{02}+\frac23\mL_{22} \right)\; ,\n\\
    \partial_w \mL_{02}
    =& -\frac{1}{w} \left[ \frac53 \mL_{02}+\frac43 \mL_{22}-\frac13 \left(1-e^{-w/2}\right)\mL_{04,\text{eq}}\right]- \left(\mL_{02}-\mL_{02,\text{eq}}\right)  \; ,\n\\
    \partial_w \mL_{22}
    =& -\frac{1}{w} \left[ \frac73 \mL_{22}-\frac{2}{15}\left(1-e^{-w/2}\right)\mL_{04,\text{eq}} \right]-\mL_{22} \; .
    \label{leom2}
\end{align}

\section{Polarization at the free-streaming fixed point}

\label{freepolapp}

 At the stable free-streaming fixed point, we can use the relations \eqref{stabfprel} to express all moments $\mG_{n\ell0}^k$ terms of $\mG^k_{\ell\ell0}$. Then \eq\eqref{pi1stterm} simplifies to
\begin{align}
 2\int dp\, p^2 E_p A^k_{n\ell}&= \Pc_n^\ell(0) \int_{p\ms} \ms^k e^{i\ell\phi} f  = \frac{\Pc_n^\ell(0)}{\Pc_\ell^\ell(0)} \int_{p\ms} \ms^k \Pc_\ell^\ell(\cos\theta) e^{i\ell\phi}= \frac{\Pc_n^\ell(0)}{\Pc_\ell^\ell(0)}  \mG^k_{\ell\ell0} \; ,
\end{align}
and \eq\eqref{pixlong} becomes
\begin{align}
   -2 \int dp\, p^2 E_p \frac{(\mathbf{A}_{n\ell}\cdot \p)p^x}{E_p(m+E_p)}&= -\int_{p\ms} \frac{E_p-m}{E_p}  \left[\ms^x  \frac14\left(e^{i(\ell+2)\phi}+2e^{i\ell\phi}+e^{i(\ell-2)\phi}\right)+\ms^y  \frac14\left(e^{i(\ell+2)\phi}-e^{i(\ell-2)\phi}\right)\right]  \Pc_n^\ell(0) f\n\\
    &= \frac14\frac{\Pc_n^\ell(0)}{\Pc_{\ell+2}^{\ell+2}(0)} \left(\mI^x_{(\ell+2)(\ell+2)0}-\mG^x_{(\ell+2)(\ell+2)0}+\mI^y_{(\ell+2)(\ell+2)0}-\mG^y_{(\ell+2)(\ell+2)0}\right)\n\\
    &+\frac14\frac{\Pc_n^\ell(0)}{\Pc_{\ell-2}^{\ell-2}(0)} \left(\mI^x_{(\ell-2)(\ell-2)0}-\mG^x_{(\ell-2)(\ell-2)0}-\mI^y_{(\ell-2)(\ell+2)0}+\mG^y_{(\ell-2)(\ell-2)0}\right)\n\\
    &+\frac12 \frac{\Pc_n^\ell(0)}{\Pc_{\ell}^{\ell}(0)} \left(\mI^x_{\ell\ell0}-\mG^x_{\ell\ell0}\right)\; .
\end{align}
Inserting these equations into \eq\eqref{pistar}, we obtain the following expression for the polarization at the free-streaming fixed point,
\begin{align}
    \Pi_\star^x&= \frac{1}{2\mathcal{N}} \frac{1}{4\pi}  \left[ \frac12 \mG^x_{000}+\frac12\mI_{000}^x+\frac14 \frac{1}{\Pc_{2}^{2}(0)} \left(\mI^x_{220}-\mG^x_{220}+\mI^y_{220}-\mG^y_{220}\right)  \right]\n\\
   & +\frac{1}{2\mathcal{N}} \sum_{n=1}^\infty \sum_{\ell=-n,\ \ell\neq 0}^{n}  N_{n\ell} \mathfrak{I}_{n\ell} \Bigg[\frac12\frac{\Pc_n^\ell(0)}{\Pc_\ell^\ell(0)}  \mG^x_{\ell\ell0}+ \frac14\frac{\Pc_n^\ell(0)}{\Pc_{\ell+2}^{\ell+2}(0)} \left(\mI^x_{(\ell+2)(\ell+2)0}-\mG^x_{(\ell+2)(\ell+2)0}+\mI^y_{(\ell+2)(\ell+2)0}-\mG^y_{(\ell+2)(\ell+2)0}\right)\n\\
    &+\frac14\frac{\Pc_n^\ell(0)}{\Pc_{\ell-2}^{\ell-2}(0)} \left(\mI^x_{(\ell-2)(\ell-2)0}-\mG^x_{(\ell-2)(\ell-2)0}-\mI^y_{(\ell-2)(\ell+2)0}+\mG^y_{(\ell-2)(\ell-2)0}\right)+\frac12 \frac{\Pc_n^\ell(0)}{\Pc_{\ell}^{\ell}(0)} \mI^x_{\ell\ell0}\Bigg] e^{i\ell\phi}\n\\
    &= \frac{1}{16\pi} \frac{1}{\mathcal{N}} \left[  \mG^x_{000}+\mI_{000}^x+\frac12 k_2 \left(\mI^x_{220}-\mG^x_{220}+\mI^y_{220}-\mG^y_{220}\right)  \right]\n\\
   & +\frac{1}{4} \sum_{\ell=-\infty, \ell\neq0}^\infty  K_\ell \Bigg[k_\ell  \left(\mG^x_{\ell\ell0}+\mI^x_{\ell\ell0}\right)+ \frac12 k_{\ell+2} \left(\mI^x_{(\ell+2)(\ell+2)0}-\mG^x_{(\ell+2)(\ell+2)0}+\mI^y_{(\ell+2)(\ell+2)0}-\mG^y_{(\ell+2)(\ell+2)0}\right)\n\\
    &+\frac12k_{\ell-2} \left(\mI^x_{(\ell-2)(\ell-2)0}-\mG^x_{(\ell-2)(\ell-2)0}-\mI^y_{(\ell-2)(\ell+2)0}+\mG^y_{(\ell-2)(\ell-2)0}\right) \Bigg] e^{i\ell\phi}
\end{align}
where $\mathfrak{I}_{n\ell}$  is defined in \eq\eqref{iint} and we introduced
\begin{align}
    K_\ell&\equiv \sum_{n=|\ell|}^\infty N_{n\ell} \mathfrak{I}_{n\ell} \Pc_n^\ell(0)\; , & k_\ell&\equiv \frac{1}{\Pc_\ell^\ell(0)}\;  .
\end{align}
However, the polarization in heavy-ion collisions is measured at freeze-out, therefore the knowledge of the polarization at the free-streaming fixed point is of limited use for practical applications in this context. In principle, one could try to derive an expression for the polarization which is valid during the full evolution from free streaming to the hydrodynamic regime. However, due to the lengths of equations and the large number of unknown moments, such calculation is hardly practicable and most likely not necessary. Therefore, in the main text we choose an expansion around local equilibrium, which is justified when considering the polarization at freeze out.

\end{appendix}

\bibliography{biblio_paper_long}{}

\end{document}